\documentclass[final,3p,times,twocolumn]{elsarticle}

\usepackage[T1]{fontenc}
\usepackage[utf8]{inputenc}
\usepackage{amsmath}
\usepackage{epstopdf}
\usepackage{flushend}
\usepackage{hyperref}
\usepackage{graphicx}
\usepackage{lineno}
\usepackage{ulem}
\usepackage{miller}
\newcommand{\ie}{\textit{i.e.}}

\newcommand{\cf}{\textit{cf}.}
\newcommand{\vasp}{\textsc{Vasp}}
\newcommand{\abinitio}{\textit{ab initio}}
\newcommand{\Abinitio}{\textit{Ab initio}}
\newcommand{\insitu}{\textit{in situ}}
\newcommand{\Insitu}{\textit{In situ}}

\usepackage{xcolor}

\newcommand{\kT}{kT}

\journal{Acta Materialia}

\begin{document}

\begin{frontmatter}

\title{Mobility of screw dislocation in BCC tungsten \\
at high temperature in presence of carbon}

\author[SRMP]{Guillaume Hachet\fnref{GPM}\corref{CA}}
\ead{guillaume.hachet@univ-rouen.fr}
\author[CEMES]{Daniel Caillard}
\author[SRMP]{Lisa Ventelon}
\author[SRMP]{Emmanuel Clouet\corref{CA}}
\ead{emmanuel.clouet@cea.fr}
\cortext[CA]{Corresponding authors}

\address[SRMP]{Université Paris-Saclay, CEA, Service de Recherches de Métallurgie Physique, 91191 Gif-sur-Yvette, France}
\address[CEMES]{CEMES-CNRS, 29 rue Jeanne Marvig, BP94347, 31055 Toulouse, France}
\fntext[GPM]{Present address: Groupe de Physique des Matériaux, Normandie University, UNIROUEN, INSA Rouen, CNRS, 76000 Rouen, France}

\begin{abstract}
The interplay of screw dislocations with carbon atoms is investigated in tungsten at high temperature using \textit{in situ} straining experiments in a transmission electron microscope (TEM) and through \textit{ab initio} calculations. 
When the temperature is high enough to activate carbon diffusion, above 1373\,K, carbon segregates in the core of screw dislocations and modifies their mobility, even for a carbon concentration as low as 1\,appm. 
TEM observations reveal the reappearance of a Peierls mechanism at these high temperatures, with screw dislocations gliding viscously through nucleation and propagation of kink-pairs.
The mobility of screw dislocations saturated with carbon atoms
is then investigated with \textit{ab initio} calculations to determine kink-pair formation, nucleation and migration energies.
These energies are used in kinetic Monte-Carlo simulations and in an analytical model to obtain the velocity of screw dislocations as a function 
of the temperature, the applied stress and the dislocation length.
The obtained mobility law parametrised on \abinitio{} calculations compares well with experiments.
\end{abstract}


\end{frontmatter}

%\linenumbers

\section{Introduction}
\label{S1}

At low temperatures, motion of screw dislocations with $1/2\,\hkl<111>$ Burgers vector
controls the development of plasticity in body-centred cubic (BCC) metals like tungsten.
These dislocations glide viscously one Peierls valley at a time by the nucleation and propagation of kink-pairs \cite{Hirth1982,Caillard2003}. 
As this Peierls mechanism is the rate controlling mechanism, gliding dislocations straighten along their screw orientation.
But, as the temperature increases, the lattice friction opposing glide of screw dislocations becomes less effective.
Above a critical temperature, screw dislocations can glide as easily as other orientations  
and plasticity proceeds by free glide of dislocations between obstacles.  
This critical temperature where the Peierls mechanism becomes athermal is around 600\,K in tungsten \cite{Brunner2000a,Brunner2010,Caillard2018}
(homologous temperature $0.16\,T_{\rm m}$, with $T_{\rm m}$ the melting temperature).

Recent tensile test experiments performed \insitu{} in a transmission electron microscope (TEM) on BCC iron of different purity \cite{Caillard2015,Caillard2016} have challenged this general understanding
of plasticity evolution with temperature in BCC metals. 
While the Peierls mechanism disappears at 300\,K in iron (homologous temperature $0.17\,T_{\rm m}$), these experiments have shown the reappearance of a lattice friction on screw dislocations at high temperature.
Dislocations lock in their pure screw orientation at 500\,K and straight screw dislocations move again viscously above this temperature.
At this transition temperature, the rate of carbon diffusion becomes comparable to the motion of mobile dislocations and, although the high-purity iron samples contain only 1 and 20\,appm of carbon, the reappearance of a Peierls mechanism at high temperature appears to be driven by the interaction of screw dislocations with carbon impurities.
\Abinitio{} calculations have confirmed this scenario by showing that the attraction between screw dislocations and carbon atoms is strong enough to lead to carbon segregation on screw dislocations, which thus glide with their carbon atmosphere, and that, even at such low nominal concentrations, the system contains enough carbon to fully saturate the dislocation cores \cite{Ventelon2015}.

\Abinitio{} calculations have shown the same strong attraction between C atoms 
and screw dislocations in tungsten \cite{Luthi2017,Bakaev2019,Hachet2020c} 
and thermodynamics predicts that screw dislocations remain fully saturated by C atoms up to 2500\,K \cite{Hachet2020c}.
A reappearance of the Peierls mechanism is therefore expected also in tungsten, 
but at a higher temperature than in iron because of the higher activation energy for C diffusion 
in tungsten than in iron, respectively 2.32 and 0.83\,eV \cite{LeClaire1990}.
Observation of dynamic strain ageing around 900\,K in tungsten ``heavy metal'' with a high carbon content \cite{Kumar1996} is another indication of a strong attraction between carbon and dislocations,  possibly in their screw orientation.
Although \insitu{} TEM straining experiments have been already performed in pure tungsten \cite{Caillard2018}, no observation of dislocation motion exists in tungsten above 600\,K.
But this temperature is too low to allow for fast enough carbon diffusion compared to dislocation motion, 
thus preventing the carbon atoms to reach thermodynamic equilibrium 
and to segregate on screw dislocations as predicted by \abinitio{} calculations \cite{Hachet2020c}.

In the present article, we perform some new \insitu{} TEM straining experiments
in tungsten at higher temperatures than in previous experiments \cite{Caillard2018}, up to 1550\,K, to confirm the reappearance of a lattice friction opposing glide of screw dislocation at high temperature.
The mobility of screw dislocations in this temperature range where C atoms are mobile and segregate 
in the dislocation cores is then analysed with the help of \abinitio{} calculations.
Formation, nucleation and migration energies of a kink-pair on a screw dislocation saturated with C atoms
are obtained from these calculations
and are used then in various kinetic models to obtain the dislocation velocity as a function of the temperature 
and of the applied stress.

\section{TEM \textit{in situ} straining experiments}
\label{S2}

\subsection{Experimental details}
\label{S21}

\textit{In situ} straining experiments are carried out in a JEOL 2010HC TEM operating at 200\,kV, with a locally developed high-temperature straining holder working up to 1573\,K, which requires an input power of 30\,W.
It is a new version of the holder described in Ref. \cite{Couret1993}.
The foil temperature has been calibrated by observing the \insitu{} melting of aluminium (933\,K), silver (1235\,K), copper (1358\,K) and silicon (1687\,K).
Microsamples are cut in the pure tungsten single crystal already used in similar experiments at lower temperatures \cite{Caillard2018}.
It is the same material as previously studied by Brunner and co-workers in Stuttgart \cite{Brunner2000a}.
It contains less than 1\,appm of oxygen, carbon, nitrogen, and silicon, and less than 0.1\,appm of other elements.
Microsamples are rectangles cut in different planes and along different directions by spark cutting.
They are mechanically polished to 10\,$\mu$m thick, and electro-polished until the formation of a thin edged hole at their centre.
They are subsequently glued on specially designed rings using a high-temperature cement, and fixed in the straining holder.
Videos are recorded at a speed of 25 images per second using a Megaview III camera, and analysed frame by frame.
Crystal orientation, Burgers vectors, slip planes and line directions of moving dislocations are easily determined in three dimensions via observations under different tilt angles and diffracting conditions.
Three samples have been successfully strained at increasing or decreasing temperatures.
They are denoted
\begin{itemize}
	\item{S1}: $\hkl(0-13)$  foil plane and $\hkl[-831]$ tensile axis,
	\item{S2}: $\hkl(100)$  foil plane and $\hkl[032]$  tensile axis,
	\item{S3}: $\hkl(-313)$ foil plane and $\hkl[101]$  tensile axis.
\end{itemize}
Videos related to these experiments can be found as supplementary materials (https://doi.org/10.1016/j.actamat.2021.117440).

\subsection{Results}
\label{S22}

All three samples exhibit the same behaviour, namely an FCC-like motion of curved dislocations below 1373\,K, and a viscous motion of straight screw dislocations above.

\begin{figure}[b!]
\centering
\includegraphics[width=0.99\linewidth]{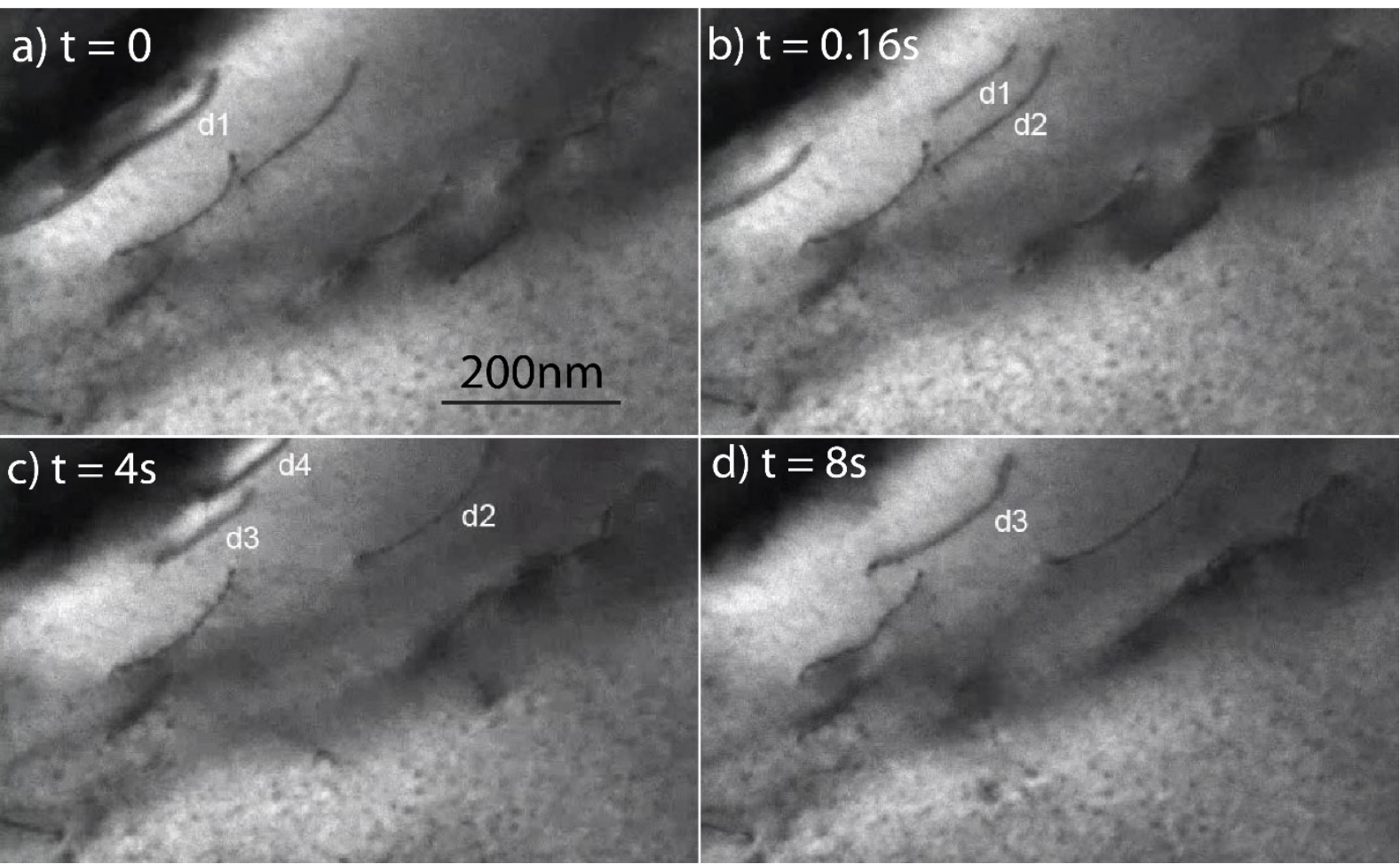}
\caption{Dislocation motion observed at 873\,K in S1 sample:
$1/2\,\hkl<111>$ dislocations are gliding in the \hkl(1-10) plane.}
\label{fig:WC_S1_600C}
\end{figure}

Fig. \ref{fig:WC_S1_600C} illustrates the low-temperature behaviour, here at 873\,K in sample S1. 
Four near-screw dislocations with Burgers vector $1/2\,\hkl[-111]$ glide in the $\hkl(101)$ plane at variable velocities, 
mostly determined by stop at and escape from obstacles which cannot be identified but may be fixed clusters of immobile solute atoms.
The instantaneous velocity between immobile positions is too high to be precisely measured.
Since displacement of more than 100\,nm are observed within less than time resolution of the video (1/50\,s), this velocity at 873\,K is larger than 5\,$\mu$m\,s$^{-1}$.
This behaviour with no clear specific characteristic feature is similar to what can be frequently observed at low
temperature in FCC metals. 

\begin{figure}[bth!]
\centering
\includegraphics[width=0.70\linewidth]{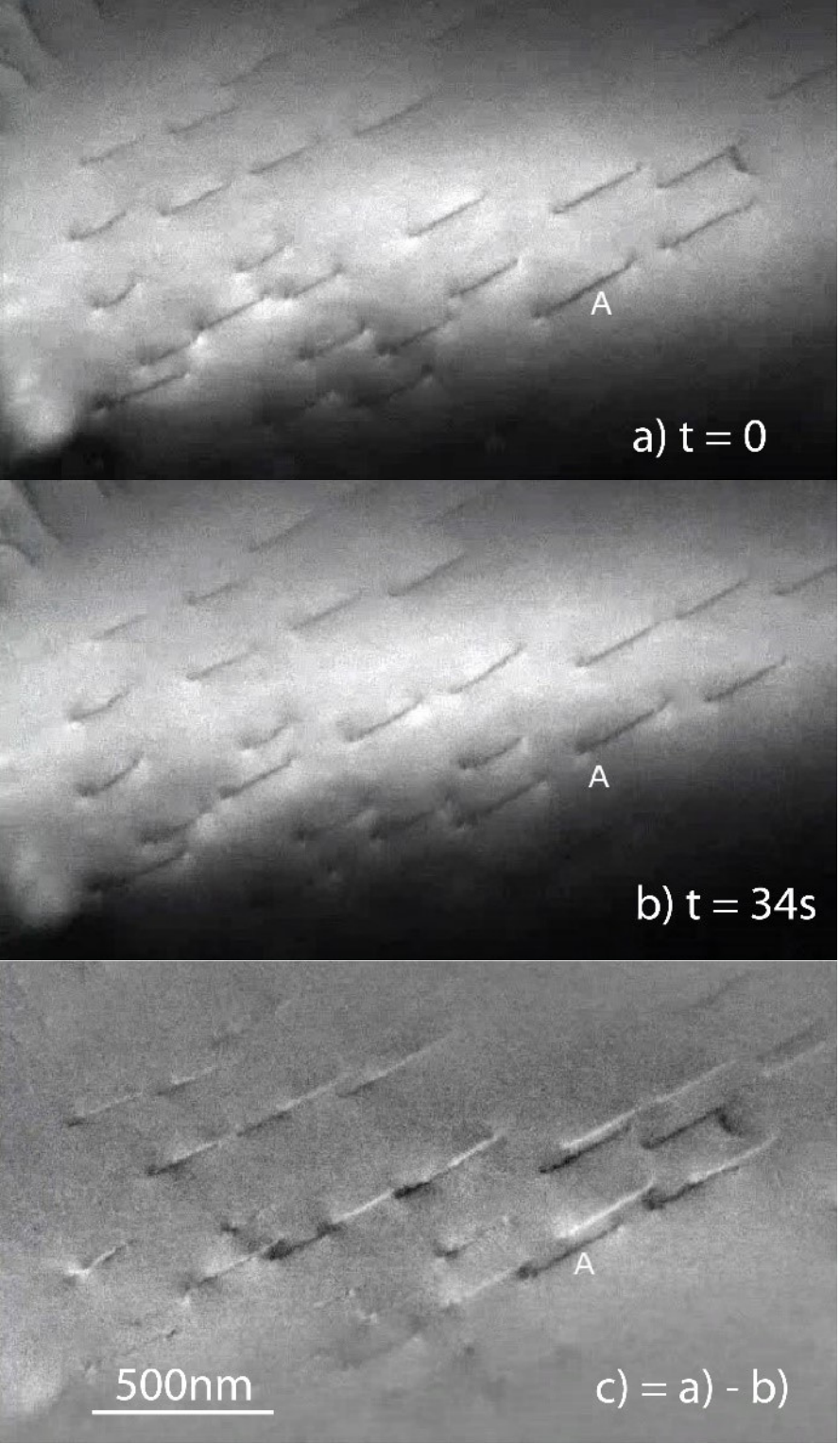}
\caption{Dislocation motion observed at 1423\,K in sample S1.
A is a fixed point and (c) is the difference between images (a) and (b).}
\label{fig:WC_S1_1150C}
\end{figure}

\begin{figure}[bth!]
\centering
\includegraphics[width=0.70\linewidth]{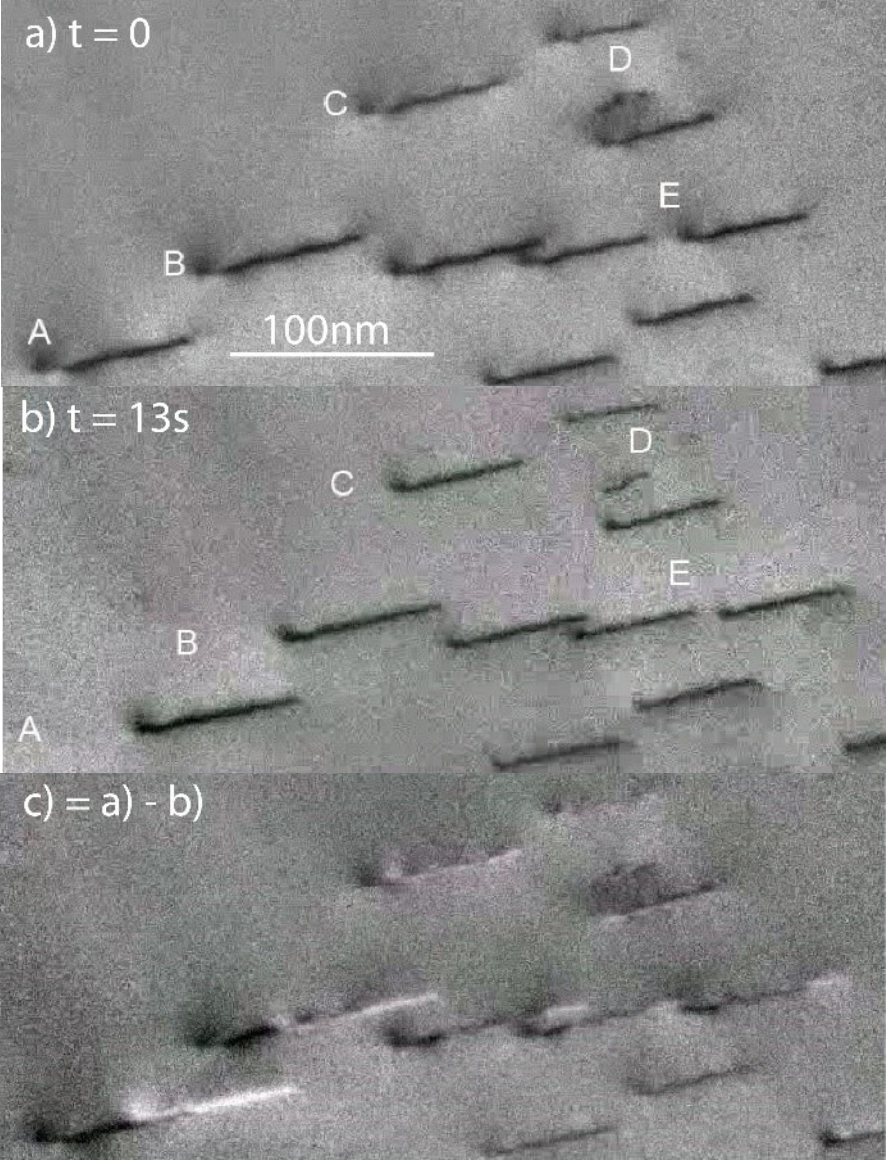}
\caption{Viscous glide of straight screw dislocations in sample S2 strained at 1423\,K.
A-E are fixed points and (c) is the difference between images (a) and (b). 
See video as supplementary material.}
\label{fig:WC_S2_1250C}
\end{figure}

The same S1 sample strained at 1423\,K is shown in Fig. \ref{fig:WC_S1_1150C}.
Dislocations with Burgers vector $1/2\,\hkl[-111]$ straighten along their screw orientation.
The screw character of the dislocations has been checked in TEM using different tilt angles.
These dislocations glide slowly and viscously in the edge-on $\hkl(-1-10)$ planes, as shown in Fig. \ref{fig:WC_S1_1150C}c which is the difference between images \ref{fig:WC_S1_1150C}a and b.
The same behaviour is observed in the S2 sample strained at 1423\,K (Fig. \ref{fig:WC_S2_1250C}): straight screw dislocations with Burgers vector $1/2\,\hkl[11-1]$ glide slowly and viscously in \hkl(011) planes.
In both dynamic sequences at 1423\,K, the dislocation velocity is rather constant and of the order of 2\,nm\,s$^{-1}$, namely much steadier and slower than at 873\,K.

\begin{figure}[bt!]
\centering
\includegraphics[width=0.99\linewidth]{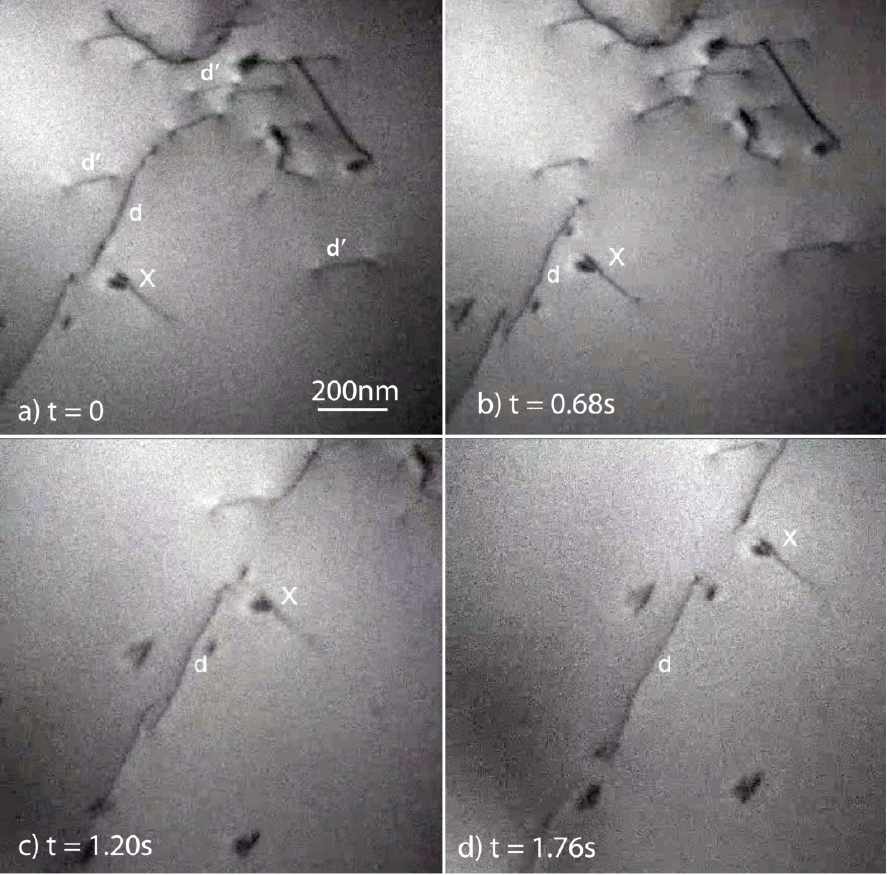}
\caption{Viscous glide of straight screw dislocations in sample S3 strained at 1413\,K.
Letter d refers to a screw dislocation with $1/2\,\hkl[-11-1]$ Burgers vector gliding downward in a \hkl(101) plane, and letters d' refer to temporarily immobile screw dislocations with $1/2\,\hkl[-111]$ Burgers vector.
X is a fixed point.
See video as supplementary material.}
\label{fig:WC_S3_1140C}
\end{figure}

\begin{figure}[bth!]
\centering
\includegraphics[width=0.70\linewidth]{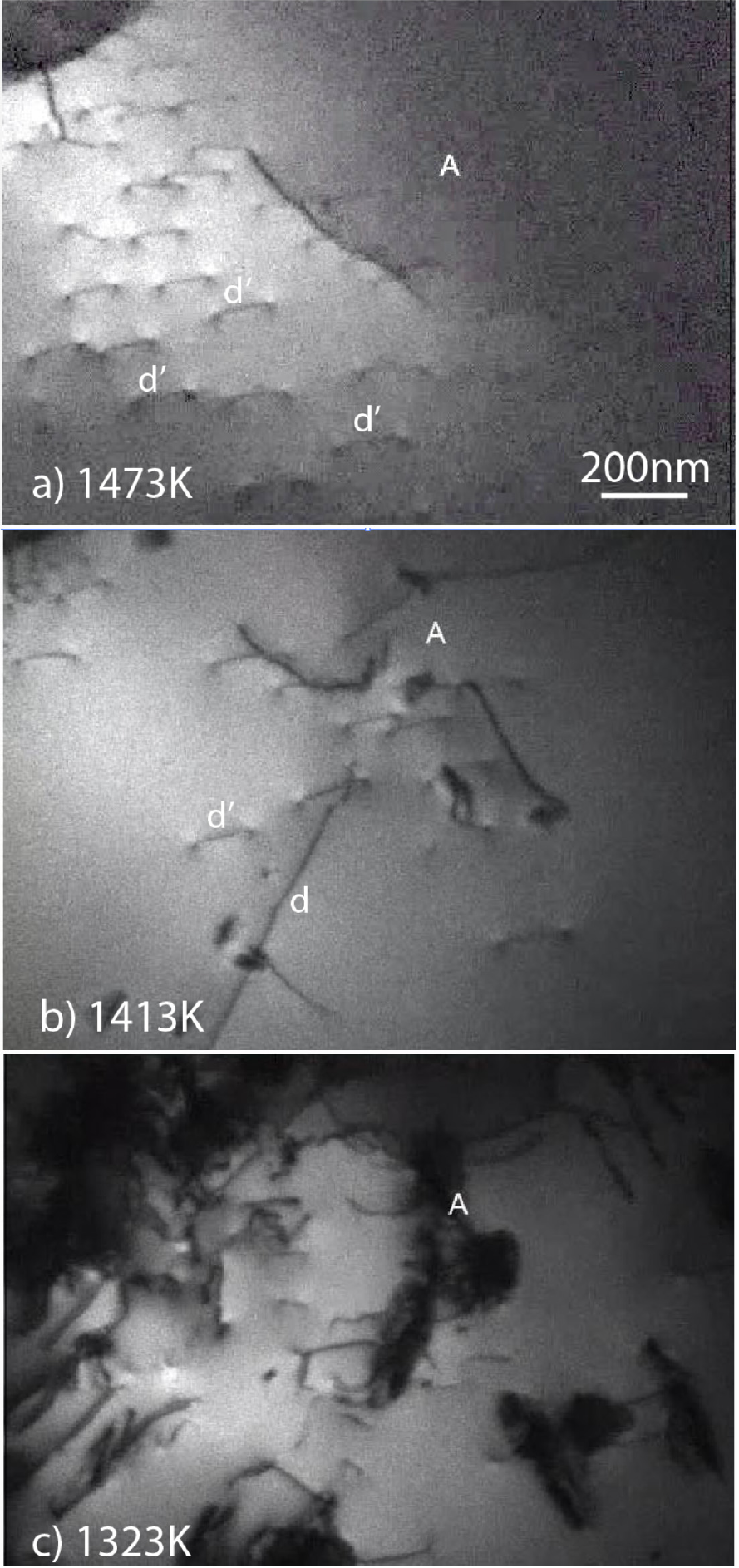}
\caption{Evolution under decreasing temperatures of dislocations noted d' in sample S3. 
The observation zone is the same as in Fig. \ref{fig:WC_S3_1140C}.
The short straight screw dislocations d' on the left of image (a) progressively disappear with decreasing temperature.
Few dislocations belonging to another slip system, like d in (b), can also be seen.
The growing precipitates near A in (c) are due to an accidental pollution of the sample surface. 
A is a fixed point.}
\label{fig:WC_S3_decreasingT}
\end{figure}

Fig. \ref{fig:WC_S3_1140C} also shows the viscous glide motion of a long screw dislocation noted d at 1413\,K in sample S3.
The $1/2\,\hkl[-11-1]$ Burgers vector is almost parallel to the foil plane and the slip plane is $\hkl(110)$.
Since this screw orientation is a priori unstable with respect to thin foil effects tending to rotate the dislocations to their shorter edge orientation perpendicular to the foil surfaces, this observation emphasises the strength of the high lattice friction maintaining the dislocations in their screw orientation at high temperature.

Another family of shorter screw segments with Burgers vector $1/2\,\hkl[-111]$ and noted d' in Fig. \ref{fig:WC_S3_1140C} is also visible.
The evolution of these shorter screw segments upon straining at decreasing temperatures is shown in Fig. \ref{fig:WC_S3_decreasingT}.
The screw dislocation segments d' visible at 1473 and 1413\,K, are no more present at 1323\,K.
This confirms that there is a transition at around 1373\,K between two different mechanisms controlling dislocation mobility.

\subsection{Discussion}

We observe here in tungsten with a low concentration (less than 1 appm) of carbon, nitrogen, oxygen and silicon the same dislocation behaviour as in ultra-pure iron containing the same low amount of the same elements \cite{Caillard2015,Caillard2016}, 
but at higher temperatures in tungsten (1400\,K) than in iron (500\,K). 
It has been shown in iron, by varying the carbon concentration, that this dislocation behaviour is controlled by carbon:
such a carbon content is responsible for the slow and viscous motion of straight screw dislocations at high temperatures 
where the dislocation glide motion becomes controlled by carbon diffusion.
\Abinitio{} calculations show the same strong interaction between carbon atoms and screw dislocations leading to carbon segregation in the dislocation core in both metals \cite{Ventelon2015,Luthi2019,Hachet2020c}, 
with a complete decoration of all dislocations provided their density remains lower than $10^{12}$\,m$^{-2}$.
\Abinitio{} results thus indicate that carbon can lead to the same viscous glide of screw dislocations in tungsten
at high temperature. 
But one cannot exclude that nitrogen, oxygen and silicon may also play a role at the investigated temperatures.

\begin{table}[t]
	\centering
	\caption{Temperature $T^{\rm act}$ (in K) at which the Peierls mechanism reappears in tungsten and iron
	and activation energy $E^{\rm diff}_{\rm X}$ (in eV) for solute diffusions 
	(values are from \cite{LeClaire1990} for C, N and O and from \cite{Sinha1971,Batz1952} for Si diffusion).
	The ratio $r$ between corresponding values in both metals is given in the last line.}
	\label{tblEaCNO}
\begin{tabular}{cccccc}
\hline
\  & $T^{\rm act}$ & $E^{\rm diff}_{\rm C}$  & $E^{\rm diff}_{\rm N}$  & $E^{\rm diff}_{\rm O}$ & $E^{\rm diff}_{\rm Si}$ \\
\hline
W   & 1400  & 2.32 & 1.55 & 1.04 & 0.40 \\
Fe  & 500   & 0.83 & 0.80 & 1.73 & 0.50 \\
$r$ & 2.8  & 2.8 & 1.9 & 0.6 & 0.8 \\
\hline
\end{tabular}
\end{table}

In iron, the reappearance of the Peierls mechanism at high temperatures 
is observed at typically $T^{\rm act}=500$\,K.
In tungsten, the same high-temperature Peierls mechanism appears at about $T^{\rm act}=1400$\,K, thus at a temperature about three times higher than in iron.  
This ratio of temperatures perfectly agrees with the ratio of activation energies for carbon diffusion in both metals (Tab.\ref{tblEaCNO})
It thus confirms that this Peierls mechanism starts when the temperature becomes high enough to activate carbon diffusion, thus allowing carbon segregation on screw dislocations, as predicted by thermodynamics \cite{Ventelon2015,Luthi2019,Hachet2020c}.
Conversely, the activation energies for diffusion in tungsten of nitrogen, oxygen, and silicon are smaller than that of carbon and cannot account for a diffusion starting at the same temperature (\cf{} ratio of activation energies in tungsten and iron in Tab. \ref{tblEaCNO}). 
We thus conclude that, as in iron above 500\,K, the slow motion of rectilinear screw dislocations observed in tungsten above 1400\,K is due to the segregation of carbon.

\section{Straight dislocation saturated by carbon}
\label{S3}

For temperatures higher than 1373\,K where \insitu{} TEM straining experiments show the reappearance  of a Peierls mechanism, 
\abinitio{} calculations and thermodynamic modelling \cite{Hachet2020c} have proved that carbon segregates on screw dislocations, with a complete reconstruction of the dislocation core towards the hard core configuration and a full saturation by C atoms of the prismatic sites created by the reconstructed core, even for nominal concentrations as low as 1\,appm C.
We study here with \abinitio{} calculations the mobility of such a reconstructed screw dislocation with all its prismatic sites occupied by carbon atoms, considering first an infinite straight dislocation, before accounting for the formation and migration of kinks in the next sections.

\subsection{Computational details}
\label{S31}

\Abinitio{} calculations are performed with \vasp{} code \cite{Kresse1999},
using the same parameters and approximations as in our previous work \cite{Hachet2020c}: pseudopotentials built with the projected augmented wave (PAW) method \cite{Blochl1994,Kresse1999} using  in the valence states 5d and 6s electrons for tungsten and 2s and 2p for carbon, a kinetic energy cutoff of 400\,eV, and the Perdew-Burke-Ernzerhof functional \cite{Perdew1996} for the exchange-correlation.
For the $1\,b$ dislocation simulation cell described below ($b$ is the dislocation Burgers vector), we use a 2$\times$2$\times$16 shifted $k$-point grid to sample the Brillouin zone and an equivalent $k$-point density for larger supercells, using a Methfessel-Paxton broadening scheme with a 0.2\,eV smearing.

The \abinitio{} calculations are performed in a periodic supercell with a constant cell volume and containing a dislocation dipole leading to a quadrupolar periodic array of dislocations \cite{Rodney2017,Clouet2018}.
Both dislocations composing the dipole are saturated with carbon atoms in their core.
The periodicity vectors $\{\mathbf{p}_1,\mathbf{p}_2,\mathbf{p}_3\}$ of the perfect supercell are defined from the elementary vectors $\mathbf{u}_1=\hkl[-1-12]$, $\mathbf{u}_2=\hkl[1-10]$ and
$\mathbf{u}_3=1/2\,\hkl[111]$:
$\mathbf{p}_1 = 5/2\,\mathbf{u}_1-9/2\,\mathbf{u}_2$,
$\mathbf{p}_2 = 5/2\,\mathbf{u}_1+9/2\,\mathbf{u}_2$
and $\mathbf{p}_3 = \mathbf{u}_3$.
This configuration implies 135 tungsten and 2 carbon atoms per $1\,b$ layer along the $Z$ axis in the \hkl[111] direction parallel to the dislocation line.
Atomic positions are relaxed until the remaining forces are less than 10\,meV/{\AA} in all Cartesian directions for static and 20\,meV/{\AA} for NEB calculations.

\subsection{Glide and energy barriers}
\label{S32}

\begin{figure}[bth!]
\centering
\includegraphics[width=0.99\linewidth]{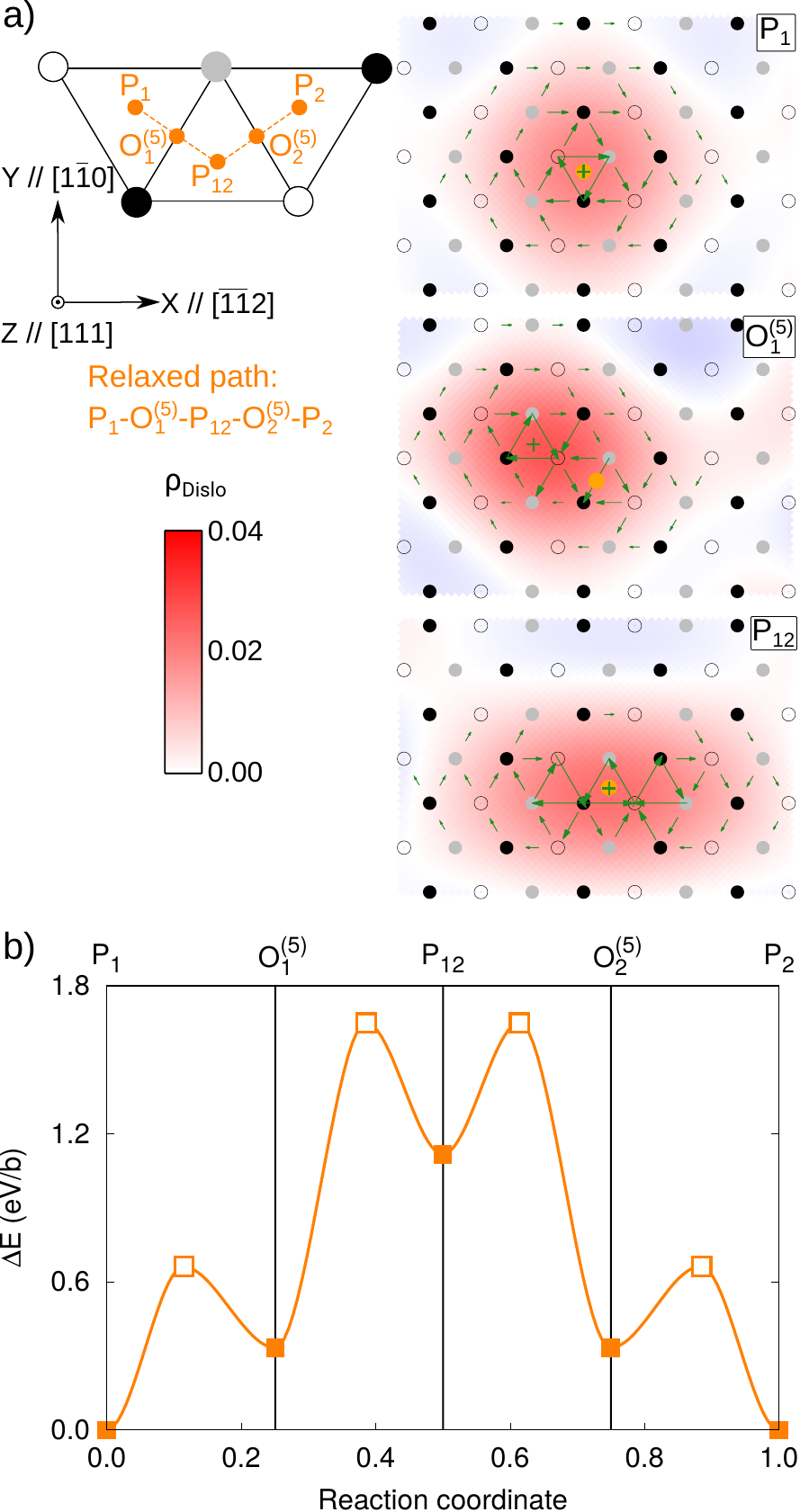}
\caption{Carbon migration from $P_1$ to $P_2$ when a straight dislocation moves from a Peierls valley to another one for $L_{\rm d}=1$\,b.
a) Sketch of the carbon displacement with the differential displacement maps of the corresponding metastable configurations obtained along the path.
In these projections perpendicular to the dislocation line, tungsten atomic columns are sketched by circles with a colour depending on their (111) plane in the original perfect crystal.
The arrows between two atomic columns are proportional to the differential displacement created by the dislocation in the \hkl{111} direction and the centre of the dislocation is marked with a green
cross.
The contour map shows the dislocation density according to the Nye tensor and the carbon atom in the vicinity of the dislocation core is shown in orange.
b) Energy variation of a straight screw dislocation saturated by carbon moving from $P_1$ to $P_2$ through the different metastable configurations.
The filled symbols are energies of relaxed configurations 
and the empty ones correspond to the energies of the transition states obtained with the climbing NEB method.}
\label{figMobStraiDislo}
\end{figure}

We first calculate the energy barrier between two stable positions, separated by one Peierls valley, of a straight dislocation saturated with carbon atoms.
The calculations are performed in a simulation cell of length $L_{\rm d} = b$ along the dislocation line, using the nudged elastic band (NEB) method \cite{Henkelman2000b} with seven intermediate images.
Both dislocations composing the dipole are displaced in the same direction
to prevent any variation of the elastic energy along the path.
Two different metastable configurations are obtained between the initial and final positions.
These configurations are then further relaxed and are presented in Fig. \ref{figMobStraiDislo}a.
The final minimum energy path is obtained with additional NEB calculations between these relaxed intermediate minima,
using one intermediate climbing image to find the saddle point.

In its initial $P_1$ and final $P_2$ positions, the dislocation is in a hard core configuration with the solute located in the prismatic interstitial site.
In the next metastable configuration along the path, the carbon atom moves to the neighbouring octahedral interstitial site and the dislocation returns to an easy core configuration, its ground state in pure tungsten.
This configuration $O^{(5)}_1$, where the carbon is in the 5$^{th}$ nearest-neighbour octahedral site of the dislocation centre, has been obtained in various BCC transition metals, including tungsten \cite{Luthi2017}.
Its energy is higher than the one of the ground state, with an energy difference $\Delta E_{P1-O5}=$ 0.33 eV/\,$b$.
Another metastable configuration, called $P_{12}$, is found in the middle of the path:
the dislocation core is spread in a \hkl{110} plane with the carbon atom in its centre.
The three atomic columns around the solute adopts a configuration similar to the hard core, thus creating a prismatic interstitial site for carbon insertion.
The energy difference between this configuration and the ground state is $\Delta E_{P1-P12}=$ 1.12 eV/\,$b$.

The whole path is then refined by performing separate NEB calculations between the different contiguous stable and metastable configurations using one intermediate climbing image  \cite{Henkelman2000b} for each portion.
The whole path appears controlled by the progressive migration of the carbon atom with the screw dislocation adopting a configuration which creates an interstitial site large enough to welcome the solute.
The two energy barriers along the path, first between configurations $P_1$ and $O_1^{(5)}$ and then between $O_1^{(5)}$ and $P_{12}$, are respectively 0.66 and 1.32\,eV/$b$ (Fig. \ref{figMobStraiDislo}b).
Such energy barriers are too high for the decorated dislocation to move as a whole from one Peierls valley to the other.
Thermal activation appears necessary to allow for dislocation glide through nucleation and migration of kink-pairs.

\section{Kink-pair formation and migration}
\label{S4}

We now use \abinitio{} calculations to determine the structure of the kinks
on the screw dislocations saturated by C atoms, and obtain the corresponding formation and migration energies, taking advantage of similar calculations performed in Fe-C system \cite{LuthiPhD}.

\subsection{Stable kink-pair configurations}
\label{S41}

Since several metastable configurations are obtained for the straight dislocation containing carbon atoms, we firstly determine the different possible stable kink-pairs which are stable.
We perform \abinitio{} calculations in a simulation cell of length $4\,b$ along the dislocation line with half the dislocation in configuration $P_1$ and the other half in configuration either $O^{(5)}_1$, $P_{12}$ or $P_2$.
Kinks built from intermediate metastable configurations $O^{(5)}_1$ or $P_{12}$ are unstable:  they relax towards a straight dislocation in both cases.
Only kinks between configurations $P_1$ and $P_2$, with the dislocation crossing one complete Peierls valley, are stable.
We will focus on these kinks in the following, using simulation cells of length either $4\,b$ or $8\,b$.

\begin{figure}[bt!]
\centering
\includegraphics[width=0.99\linewidth]{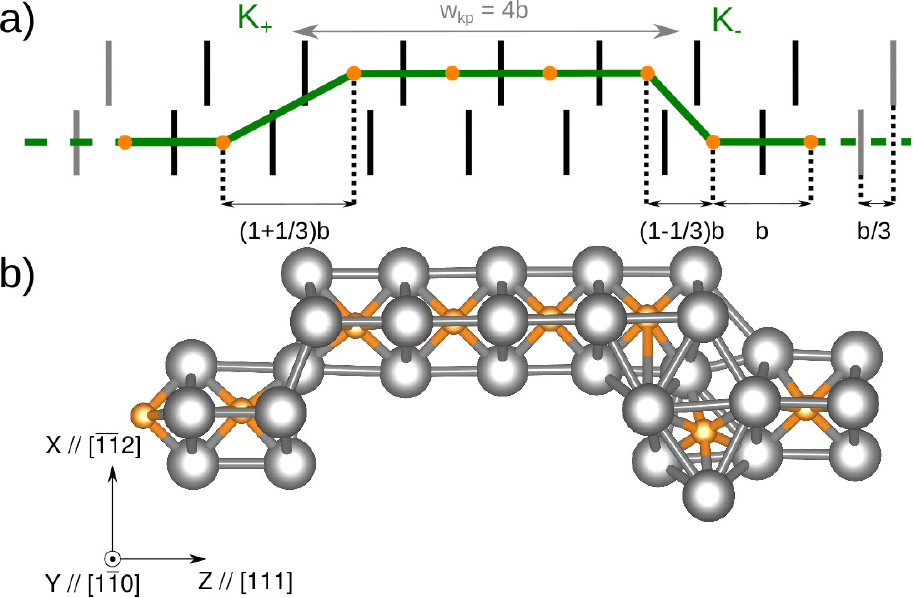}
\caption{Kink-pair with a separation distance $w_{\rm kp} = 4\,b$ on a dislocation of length $L_{\rm d}=8\,b$.
a) Sketch of the kinked dislocation saturated by carbon.
The green line represents the dislocation, the orange circles the C atoms, and the vertical segments the \hkl(111) planes.
b) Relaxed atomic structure of the kinked dislocation.
Tungsten and carbon atoms are shown respectively in grey and orange.
For tungsten, only atoms belonging to the dislocation core or first neighbours of a carbon atom are represented.
All carbon atoms are lying in a prismatic site, except the one near the $K_{-}$ kink which has been ejected in an octahedral site.}
\label{figDisloDK8b}
\end{figure}

Both kinks composing a pair are abrupt, localised in a region of only $1\,b$ width (Fig. \ref{figDisloDK8b}), contrary to kinks in pure BCC metal which spread over a with of $\sim20\,b$ \cite{Rao2001,Ventelon2009,Proville2013,Dezerald2015}.
This is a consequence of the C atoms segregated in the dislocation core which pin the dislocation in the bottom of its Peierls valley.
Because of the different atomic arrangement, both kinks are not equivalent: one kink is in tension while the other is compressed along the dislocation line due to a $b/3$ shift in the $\hkl[111]$ direction (Fig. \ref{figDisloDK8b}a).
They will be respectively called $K_{+}$ and $K_{-}$ \cite{Seeger1976,LuthiPhD}.
Besides, the carbon atom near $K_{-}$ moves from the prismatic site to the nearest octahedral site (Fig. \ref{figDisloDK8b}b) because the former interstitial site is too highly strained inside the kink.

\begin{figure}[bth]
\centering
\includegraphics[width=0.49\linewidth]{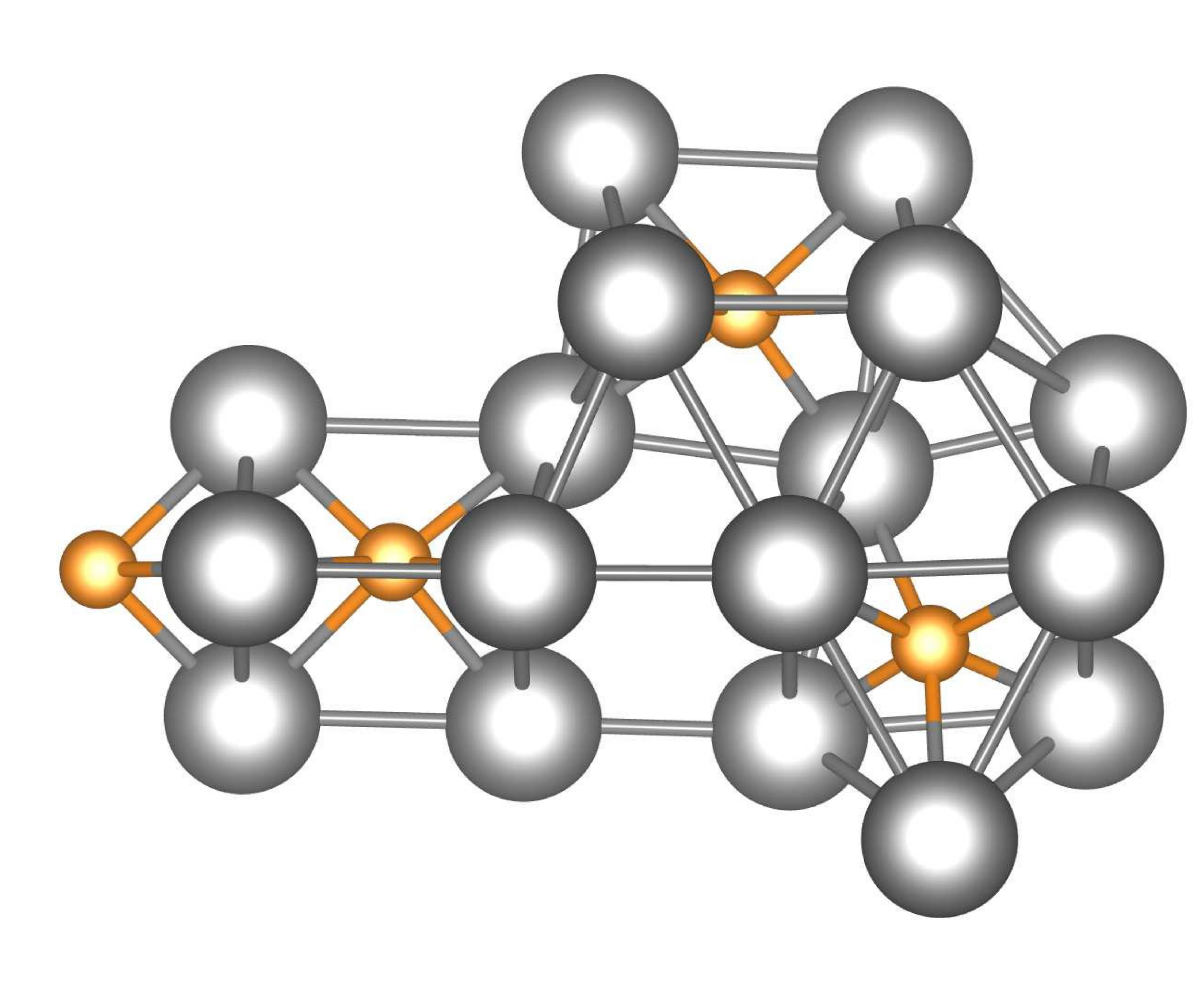}
\caption{Relaxed atomic structure of the kinked dislocation with a separation distance $w_{\rm kp} = 1\,b$ on a dislocation of length $L_{\rm d}=4\,b$.}
\label{figDisloDK1b}
\end{figure}

We also perform \abinitio{} calculations of kinked dislocations for the smallest separation distance $w_{\rm kp} = 1\,b$,
using a supercell of length $L_{\rm d}=4\,b$.
The relaxed atomic configuration is stable and presented in Fig. \ref{figDisloDK1b}.
In this elementary pair, $K_+$ and $K_-$ kinks are next to each other, 
with the carbon atom ejected from the initial prismatic site to the closest octahedral interstitial site at the $K_{-}$ kink, 
like for largest pairs.

\subsection{Formation energy}
\label{S42}

\begin{figure}[b!]
\centering
\includegraphics[width=0.99\linewidth]{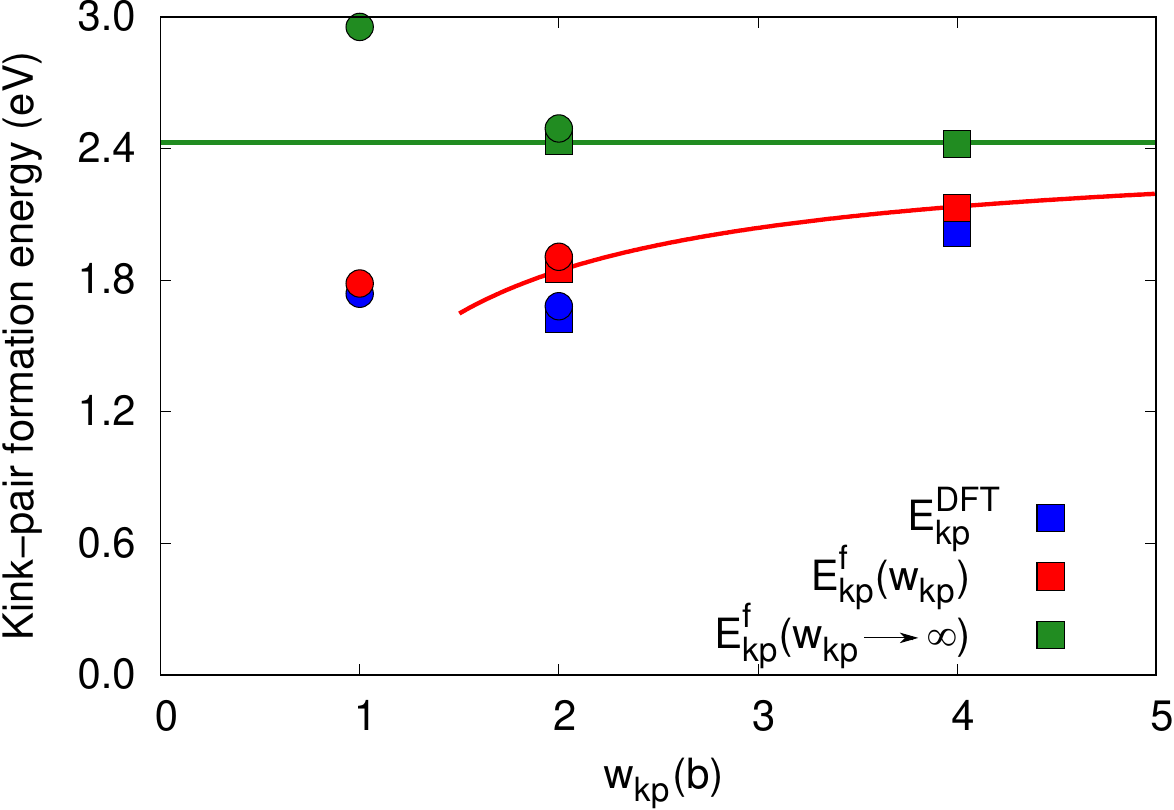}
\caption{Kink-pair formation energy determined before (blue) and after (red) elastic correction as a function of the kink-pair separation distance ($w_{\rm kp}$).
The red curve describes the formation energy of an isolated pair of kinks (Eq. \ref{eqFormEnergy}).
The energies and curves in green correspond to the formation energy of two infinitely separated single kinks.
Either one (circles) or both (squares) dislocations of the dipole are kinked.}
\label{figEform}
\end{figure}

We now determine the kink-pair formation energy $E^{\rm f}_{\rm kp}$ by using simulation cells of length $L_{\rm d}=4\,b$ and $8\,b$ containing dislocations with a kink-pair of width $w_{\rm kp}=b$ ($L_{\rm d}=4\,b$), $2\,b$ ($L_{\rm d}=4\,b$), and $4\,b$ ($L_{\rm d}=8\,b$).
The DFT formation energy $E^{\rm DFT}_{\rm kp}$ is defined as the energy difference between the same supercells containing kinked and straight dislocations.
Kink-pairs are created on only one or the two dislocations composing the dipole, but the obtained formation energy is quite insensitive to this choice (see the difference between squares and circles for $w_{\rm kp}=2\,b$ in Fig. \ref{figEform}).
On the other hand the formation energy depends on the separation width $w_{\rm kp}$ and on the length $L_{\rm d}$ of the simulation cell, because the kinks interact elastically.
Part of this interaction energy arises from the periodic boundary conditions along the dislocation lines, as the kinks interact with their periodic images.
As it will be shown below, this spurious interaction can be removed using elasticity theory to retrieve the formation energy of an isolated kink-pair on an infinite dislocation.

When the width $w_{\rm kp}$ of the kink-pair is larger than its height, \ie{} the distance between two Peierls valleys $h=a_0\sqrt{2/3}$ ($a_0=3.173$\,\AA), the elastic interaction between the two kinks of opposite sign is well approximated by \cite{Lothe1992}
\begin{equation}
	W_{\rm int}(w_{\rm kp}) = - \theta\frac{h^2}{w_{\rm kp}},
\label{eqLotheElasInt}
\end{equation}
with $\theta$ the line tension coefficient.
Assuming isotropic elasticity, a good approximation for tungsten, this parameter is equal to \cite{Lothe1992}
\begin{equation}
\theta=\frac{\mu b^2}{8 \pi}\frac{1+\nu}{1-\nu},
\label{eqLotheLineTens}
\end{equation}
with $\mu=133$\,GPa and $\nu=0.32$, respectively the shear modulus and Poisson ratio of tungsten, obtained by Voigt-Reuss-Hill average of the elastic constants $C_{11} = 497$, $C_{12} = 227$, and $C_{44} = 131$\,GPa.

The total elastic interaction energy is obtained by summing the interaction between the two kinks inside the simulation cell and also with their periodic images on the dislocation line, considering their alternating signs.
One obtains an infinite summation which can be reduced to a simple analytical expression for the two specific cases considered in our simulations:
\begin{align}
	W_{\rm int}^{\rm tot} =& -\theta \frac{h^2}{L_{\rm d}} 4 \ln{(2)} \textrm{\quad for \quad} w_{\rm kp}=\frac{L_{\rm d}}{2}, \\
	=& -\theta \frac{h^2}{L_{\rm d}} 6 \ln{(2)} \textrm{\quad for \quad} w_{\rm kp}=\frac{L_{\rm d}}{4}.
\end{align}

Removing this total interaction energy, one obtains the formation energy of two infinitely separated single kinks, $E^{\rm f}_{\rm kp}(w_{\rm kp} \rightarrow \infty)$
(Fig. \ref{figEform}).
The same formation energy equal to 2.43\,eV is obtained for $w_{\rm kp}=2\,b$ and $4\,b$, thus showing that the elastic model perfectly describes the interaction between kinks and allows obtaining results free from boundary effect and supercell size.
Even for such small separation distance between kinks, the kink-pair can be considered as formed of two well developed kinks interacting only elastically.

Inversely, for the smallest kink-pair, \ie{} for $w_{\rm kp}=b$,
the value obtained for $E_{\rm kp}^{\rm f}(w_{\rm kp}\rightarrow\infty)$
differs from the one of largest pairs.
For such a small separation distance, the kink-pair cannot be described as two isolated kinks which interact elastically
and a full atomistic description is needed.
On the other hand, the elastic interaction of the kink-pair with its periodic images can be safely neglected in a supercell of length $L_{\rm d}=4\,b$ (see difference between red and blue symbols
for $w_{\rm kp}=b$ in Fig. \ref{figEform}).

Finally, the formation energy of an isolated pair of kinks separated by a finite distance $w_{\rm kp}$ is given by
\begin{equation}
	E^{\rm f}_{\rm kp} (w_{\rm kp}) = E^{\rm f}_{\rm kp} (w_{\rm kp} \rightarrow \infty) - \theta \frac{h^2}{w_{\rm kp}}.
\label{eqFormEnergy}
\end{equation}
This equation is valid for any distance larger than $2\,b$ up to $\infty$.
For $w_{\rm kp}=1b$, this analytical expression underestimates the formation energy (Fig. \ref{figEform}) and the value given by \abinitio{} calculations needs to be used.

\subsection{Kink migrations}
\label{S43}

\begin{figure}[b!]
\centering
\includegraphics[width=0.99\linewidth]{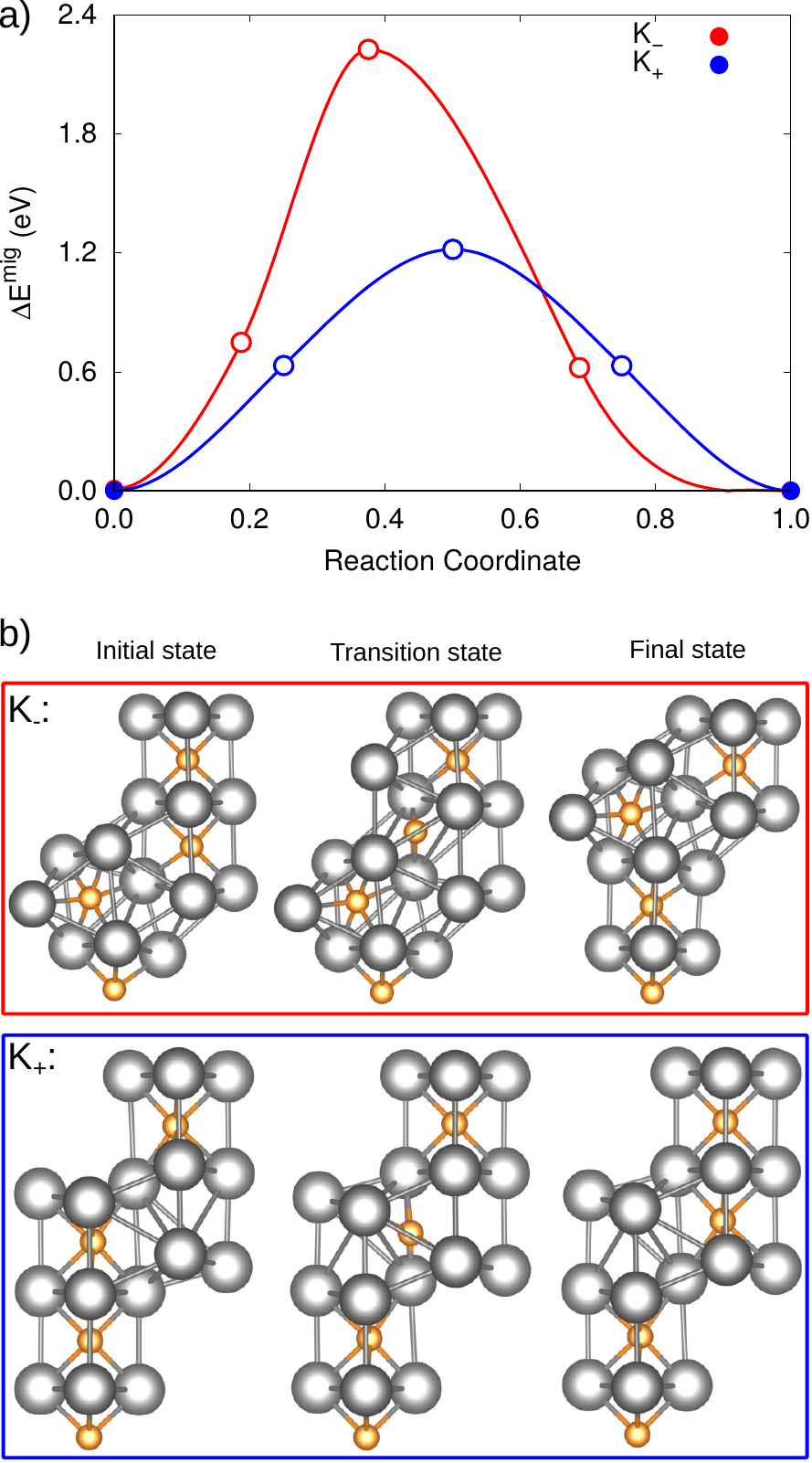}
\caption{Migration of $K_{+}$ and $K_{-}$ kinks obtained with NEB calculations.
a) Energy barriers.
b) Relaxed atomic configurations along the path: initial, saddle point, and final states.}
\label{figNEBMig}
\end{figure}

Then, we determine the migration energies of both $K_{+}$ and $K_{-}$ kinks by performing NEB calculations on simulation cells of height close to $4\,b$.
For these simulations, only one kink, either $K_{+}$ or $K_{-}$, is inserted on each dislocation.
As a consequence, the periodicity vector $\mathbf{p_3}$ along the dislocation line needs to be tilted,
with $\mathbf{p_3}=(3n+1)/6\,\hkl[111] - 1/3\hkl[-1-12]$ for $K_{+}$
and $\mathbf{p_3}=(3n-1)/6\,\hkl[111] + 1/3\hkl[-1-12]$ for $K_{-}$ \cite{Ventelon2009}.
For $n=4$, this leads to simulation cells containing either 585 or 495 tungsten atoms for $K_{+}$ and $K_{-}$, respectively.
Although the two dislocations forming the dipole are kinked, only one kink,
and the corresponding C atom, is moved during the NEB calculations.
A threshold of 10\,meV/{\AA} on atomic forces is used for these NEB calculations.

The obtained energy barriers are presented in Fig. \ref{figNEBMig}a.
The heights of these barriers are equal to $E^{\rm mig}_{+}=1.22$ and $E^{\rm mig}_{-}=2.22$\,eV respectively for $K_{+}$ and $K_{-}$.
As the kink is moving along the line with its periodic images, there is no variation of the interaction energy 
between the kink and its images during the migration path.  Besides, this constant interaction energy is small, 
around 0.1\,eV per kink as predicted by elasticity (see difference between red and blue symbols in Fig. \ref{figEform}
for a kink pair of width $w_{\rm kp}=2\,b$). 
The energy barriers given by these NEB calculations can therefore be considered as the ones for the migration of an isolated kink.
The barrier is much higher for the $K_{-}$ kink as the migration of this kink involves the jump of two carbon atoms (Fig. \ref{figNEBMig}b),
one carbon atom from a prismatic to an octahedral site, and another one from an octahedral to a prismatic site.
On the other hand, for $K_{+}$ migration, only one carbon atom is jumping, directly between two prismatic sites.

\subsection{Kink-pair nucleation}
\label{S44}

\begin{figure}[bth!]
\centering
\includegraphics[width=0.99\linewidth]{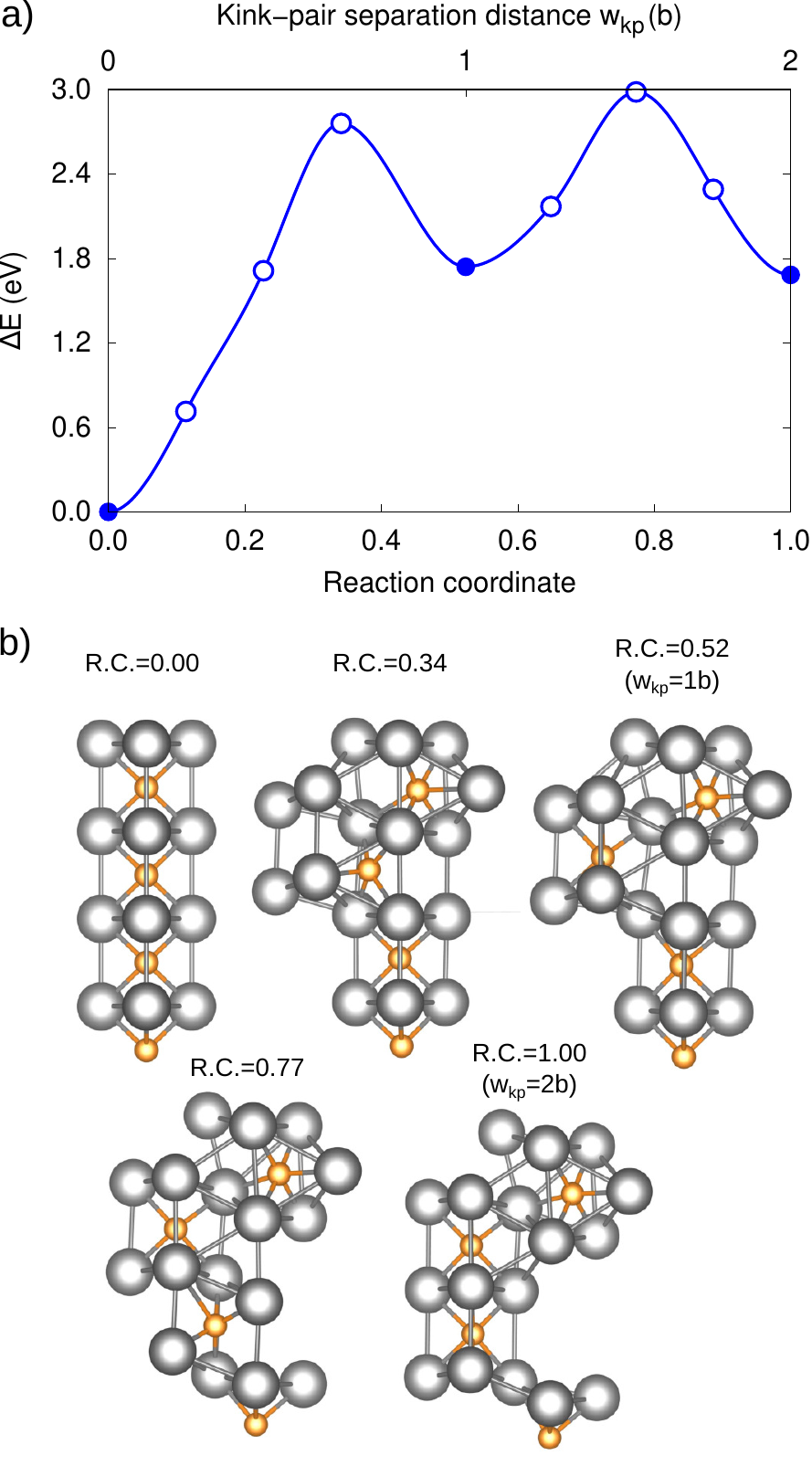}
\caption{Nucleation on a straight dislocation of a kink-pair with a separation distance $w_{\rm kp}=2\,b$.
a) Energy barrier.
b) Relaxed atomic configurations along the path which goes through an intermediate metastable configuration for the reaction coordinate 0.52 corresponding to a $1\,b$ kink-pair.}
\label{figNEBNuc}
\end{figure}

Finally, we determine the energy barrier for the nucleation of a pair of kinks on a straight dislocation saturated by carbon.
We consider in the NEB calculation a final state with a kink-pair separated by a distance $w_{\rm kp}=2\,b$, as this corresponds to a separation distance for which kinks are fully developed, with a formation energy well described by elasticity (Eq. \ref{eqFormEnergy}).
The length of the simulation cell is $L_{\rm d}=4\,b$ and the kink-pair is created on only one dislocation of the dipole.

The results of these calculations are presented in Fig \ref{figNEBNuc}a.
The transition path goes through two energy barriers,
with a metastable intermediate state corresponding to a pair of kinks with a minimal separation distance
$w_{\rm kp}=b$.
The first energy barrier, $E^{\rm nuc}_{1b}=2.76$\,eV, for the nucleation of the $1\,b$ kink-pair is slightly lower than the height of the second transition state, 2.97\,eV corresponding to the migration of the $K_{+}$ kink.
This second barrier is almost equal to the sum of the formation energy of the $1\,b$ kink-pair (1.73\,eV)
and of the migration energy of an isolated $K_{+}$ kink, $E^{mig}_{+}=1.22$\,eV.
Although the separation distance between the two kinks is small,
their migration energy is already very close to the one of an isolated kink.

\section{Mobility of carbon saturated screw dislocation}
\label{S5}

We now determine the velocity of a screw dislocation saturated by carbon atoms 
from the kink-pair formation and migration energies determined in the previous section. 
Dislocation velocities are first obtained with kinetic Monte Carlo (kMC) simulations, 
neglecting interactions between kinks. 
Results are used to demonstrate the accuracy of analytical
expressions describing the dependence of the dislocation velocity
with the temperature, the applied stress, and the dislocation length. 
The analytical model is finally enriched to take full account of the energy profile 
of the kinked dislocation, including its variation with the width of the kink-pair, 
thus giving a quantitative description of dislocation mobility.

\subsection{Energy profile of a kinked dislocation}
\label{sec:EKinkedDislo}

\begin{figure}[b!]
\centering
\includegraphics[width=0.99\linewidth]{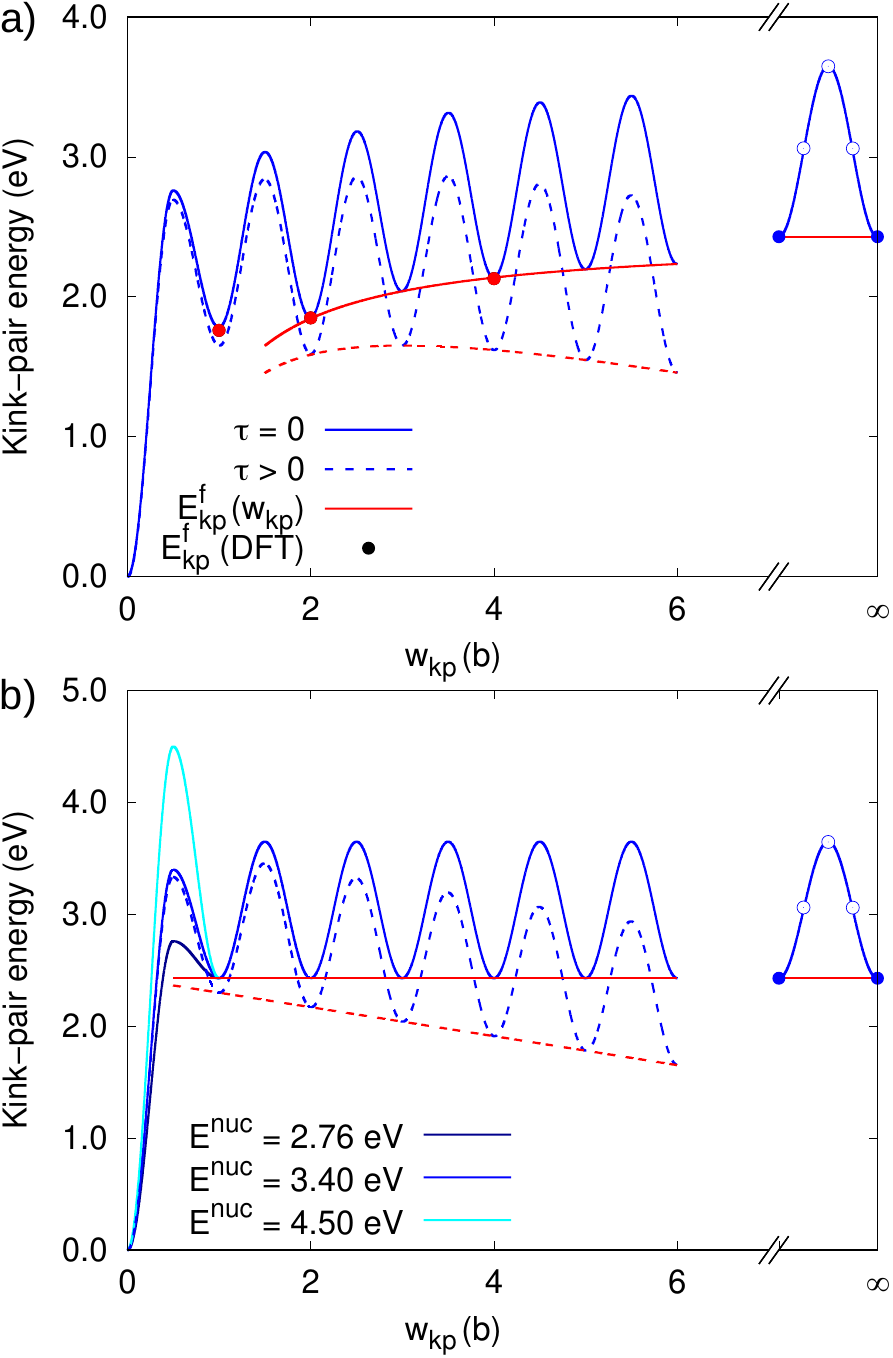}
\caption{Energy profile describing the evolution of a screw dislocation saturated with carbon atoms 
gliding by nucleation and propagation of a pair of kinks with a width $w_{\rm kp}$. 
The energy profile in (a) has been deduced from the \abinitio{} calculations described in section \ref{S4}, whereas the simplified profiles in (b) are the ones used in the kMC simulations with three different values of the nucleation energy $E^{\rm nuc}$.
For both figures, the dashed lines correspond to the evolution with an applied stress $\tau=1$\,GPa.}
\label{figKPMod}
\end{figure}

>From the previous DFT results, we can describe the energy landscape of 
a dislocation saturated by carbon atoms containing a pair of kinks separated by a distance $w_{\rm kp}$, 
from $w_{\rm kp}=0$ to $w_{\rm kp} \rightarrow \infty$ (Fig. \ref{figKPMod}a).
When $w_{\rm kp}\leq 2\,b$, the kink-pair energy follows the DFT results from section \ref{S44} 
with two successive energy barriers equal to 2.76 and 2.97\,eV 
and an intermediate stable configuration for $w_{\rm kp}=1\,b$ for which the formation energy is 1.73\,eV.
For larger width, the formation energy of the kink-pair is described by the elastic interaction model (Eq. \ref{eqFormEnergy}) and the activation energy for transition between widths $w_{\rm kp}=n\,b$ and $(n+1)\,b$ is taken equal to the kink migration energy corrected for the energy difference between the initial and final configurations, $E^{\rm act}_{n \to n+1} = E^{\rm mig} + ( E^{\rm f}_{n+1} - E^{\rm f}_{\rm n} )/2$, with $E^{\rm f}_{n}=E^{\rm f}_{\rm kp}(w_{\rm kp}=n\,b)$.
Both migration of $K_{+}$ and $K_{-}$ should theoretically be considered, but kMC simulations and the analytical model described below show that only $K_{+}$ migration is active below 3000\,K, the maximal temperature considered in this study. 
Finally, when an external stress $\tau$ is applied, a contribution $-\tau\,b\,h\,w_{\rm kp}$ corresponding to the work of the Peach-Koehler forces is added to the kink-pair formation energy (\cf{} dashed lines in Fig. \ref{figKPMod}a).

The size dependence of the kink-pair formation energy adds some complexity 
to the kinetic Monte Carlo simulations, leading to time consuming simulations \cite{Cai1999,Bulatov2006}.
We therefore consider first an idealised energy landscape (Fig. \ref{figKPMod}b) where the formation energy is constant 
and equal to the value for two non interacting kinks, $E^{\rm f}_{\rm kp}=E^{\rm f}_{\rm kp}(w_{\rm kp}\to\infty)=2.43$\,eV. 
Kink migration energies are also taken constant, with $E^{\rm mig}_{+}=1.22$\,eV 
and $E^{\rm mig}_{-}=2.22$\,eV for kink width $w_{\rm kp}\geq b$.
Only the first energy barrier $E^{\rm nuc}$
corresponding to the nucleation of a $1\,b$ kink-pair takes a different value.
As it is not possible, with such an idealised energy landscape, 
to reproduce the real energy barriers for both kink-pair nucleation and annihilation,
we test different values for this first barrier, 
to  check the ability of the analytical model to capture the different situations.
Once this analytical model has been validated, 
the real energy landscape with the formation energy of interacting kinks
will be considered.

\subsection{Kinetic Monte Carlo simulations}
\label{sec:KMC}

The kMC simulations follow the work of Bulatov and Cai \cite{Bulatov2006}, where a dislocation of length $L_{\rm d}$ is discretised with nodes separated by $1\,b$.
Each node can move either in the direction of the applied stress (forward) or in opposite direction (backward).
Elementary dislocation segments are moving with their carbon atom segregated in the core. 
Each node motion can thus be seen as a jump of a carbon atom, with the dislocation remaining pinned on the solute.
Backward/forward motions of nodes lead to different events depending on the positions of the neighbouring nodes: 
nucleation/annihilation of a $1\,b$ kink-pair or kink migration changing the width of an already existing kink-pair.
The frequency of each event takes the following form:
\begin{equation}
	\Gamma = \nu_{\rm D} \exp{\left(-\frac{E^{\rm act} + \Delta E/2}{\kT}\right)},
	\label{eqTransRate}
\end{equation}
with $E^{\rm act}$ an activation energy and $\Delta E$ the energy variation between the initial and final states.
These energies are directly linked to the kink-pair formation and nucleation energies and kink migration energies and are presented in Tab. \ref{tblEmDE}.
The attempt frequency $\nu_{\rm D}$ is taken equal to the Debye frequency 
($\nu_{\rm D}=8.3$\,THz \cite{Kittel2004}).

\begin{table}[h]
	\centering
	\caption{Activation energy $E^{\rm act}$ and energy variation $\Delta E$
	entering in the frequency definition for the different kMC events (Eq. \ref{eqTransRate}).
	The contribution $\tau h b^2$ of the resolved shear stress $\tau$ is negative for a forward motion of the node
	and positive for backward motion.}
	\label{tblEmDE}
\begin{tabular}{lcc}
\hline
 			& $E^{\rm act}$ 			& $\Delta E$ \\
\hline
Nucleation 		& $E^{\rm nuc} - E^{\rm f}_{\rm kp}/2$ 	& $ E^{\rm f}_{\rm kp} \mp \tau h b^2$ \\
Annihilation 		& $E^{\rm nuc} - E^{\rm f}_{\rm kp}/2$ 	& $-E^{\rm f}_{\rm kp} \mp \tau h b^2$ \\
$K_{+}$ migration 	& $E^{\rm mig}_{+}$ 			& $\mp \tau h b^2$ \\
$K_{-}$ migration 	& $E^{\rm mig}_{-}$ 			& $\mp \tau h b^2$ \\
\hline
\end{tabular}
\end{table}

\begin{figure}[b!]
\centering
\includegraphics[width=0.99\linewidth]{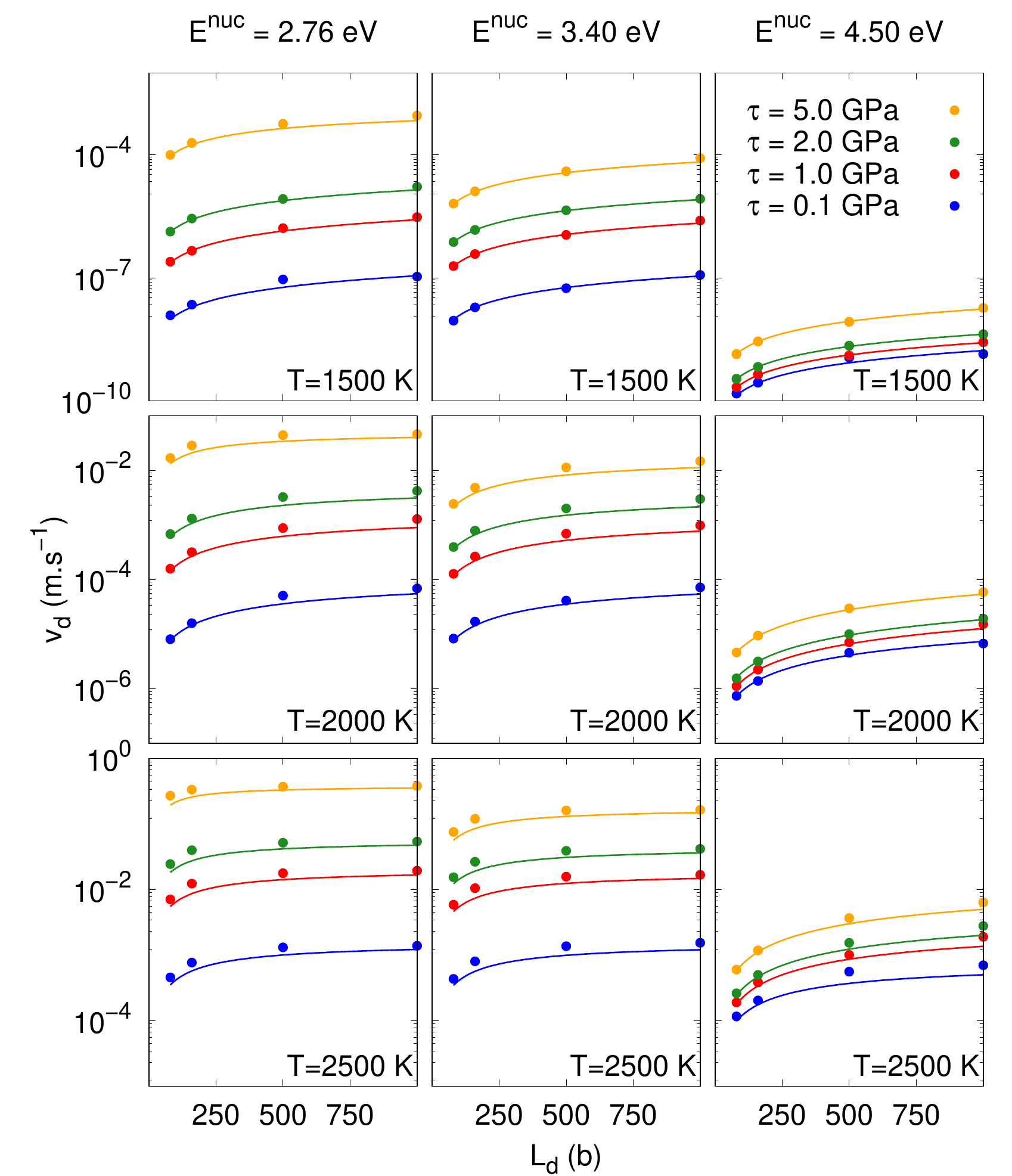}
\caption{Dislocation velocity as a function of the dislocation length for different temperatures and external stresses.
The dots correspond to results of the kMC simulations and the lines to the analytical model given by Eq. \ref{eqvD} for the simplified energy profile of Fig. \ref{figKPMod}b.}
\label{figvDSimpMod}
\end{figure}

At each time step of the kMC simulations, each node of the dislocation line can move forward or backward, the probability for each of these events being proportional to the frequency $\Gamma$ (Eq. \ref{eqTransRate}).
Following a time residence algorithm \cite{Bulatov2006}, one of these events is selected randomly and accepted or rejected according to the probability distribution. 
The time is incremented with the inverse of the sum of all events frequencies.
The dislocation velocity is computed from the time evolution of the dislocation average position. 
Simulations are performed for different values of the nucleation energy: 
$E^{\rm nuc}=2.76$, 3.40, and 4.50\,eV. 
Results of kMC simulations are shown in Fig. \ref{figvDSimpMod} for different dislocation lengths $L_{\rm d}$, applied stresses $\tau$ and temperatures $T$.
One sees strong variations of the dislocation velocity with all these parameters, which would be analysed below with the help of an analytical model predicting this velocity.

\subsection{Dislocation velocity}
\label{S52}

Following Refs. \cite{Hirth1982,Caillard2003,Kubin2013,Po2016}, the velocity of a dislocation gliding by nucleation and propagation of kinks
is given by
\begin{equation}
	v_{\rm d} = J_{\rm kp} \, h \, \frac{L_{\rm kp}}{b},
\label{eqvD}
\end{equation}
with $h$ the distance between two Peierls valleys, $J_{\rm kp}$ the nucleation rate of kink-pairs, and $L_{\rm kp}$ the average extension of a kink-pair, \ie{} the sum of the mean free paths of the two kinks composing the pair.
As shown in \ref{app:kinkNucleation}, the nucleation rate can be written
\begin{multline}
	J_{\rm kp} = 2 \nu_{\rm D} \sinh{\left(\frac{\tau hb^2}{2\kT}\right)} \exp{\left(-\frac{E^{\rm mig}_+}{\kT}\right)} \ / \\
	\left\{ \exp{\left(\frac{E^{\rm nuc} - E^{\rm mig}_+}{\kT}\right)} \left[ 1 - \exp{\left( - \frac{\tau h b^2}{\kT} \right)} \right] \right. \\ \left.
	+ f(\tau,T) \exp{ \left( \frac{E^{\rm f}_{\rm kp} - \tau h b^2}{\kT} \right)}\right\}.
\label{eqJTot}
\end{multline}
The function $f(\tau,T)$ incorporates all the variation of the kink-pair formation energy with its width. 
For the idealised energy landscape of Fig. \ref{figKPMod} where the formation energy is constant, this function is simply equal to 1.
Only migration energy of the $K_+$ kink appears in Eq. \ref{eqJTot} as it is smaller than the one for $K_{-}$.  
As a consequence $K_{-}$ migration can be neglected in the temperature range of the kMC simulations, an assumption we have checked with the general expression (Eq. \ref{eq:nucleationRate}) where migration of both kinks is considered.

The kink-pair average extension $L_{\rm kp}$ can have two different expressions, depending on the dislocation length $L_{\rm d}$.
For small dislocations, as at most one kink-pair exists on the dislocation line at a time, the kink-pair average extension is simply equal to the dislocation length, $L_{\rm kp}=L_{\rm d}$, thus leading to the length dependent regime for the dislocation velocity.
When the dislocation becomes long enough, several kinks can coexist on the line and their mean free path becomes controlled by their collisions statistics. 
In this kink collision regime, each kink travels in average a distance $L_{\rm kp}/2=v_{\rm k} \, \Delta t_{\rm kp}$ in a time interval $\Delta t_{\rm kp}$ between two collisions, with $v_{\rm k}$ the average velocity of a kink.
The kink-pair average survival time $\Delta t_{\rm kp}$ and extension $L_{\rm kp}$ are also linked to the nucleation rate by $J_{\rm kp} \, \Delta t_{\rm kp} \, L_{\rm kp}/b = 1$.
As a consequence, the average extension of a kink-pair is given in this kink collision regime by $L_{\rm kp} = \sqrt{2bv_k / J_{\rm kp}}$.
To obtain a simple expression of the kink-pair extension which is valid in both the length dependent and the kink collision regimes, we consider the geometric average \cite{Po2016}
\begin{equation}
	L_{\rm kp} = \frac{1}{ 1/L_{\rm d} + \sqrt{J_{\rm kp} / 2bv_k} },
	\label{eq:nucleation_kink_extension}
\end{equation}
which leads to a smooth interpolation between values corresponding to each regime.
As shown in \ref{app:kinkVelocity}, the kink velocity 
appearing in this equation is given by
\begin{equation}
	v_{\rm k} = b \, \nu_{\rm D}
		\exp{\left(-\frac{E^{\rm mig}_+}{\kT}\right)}
		\sinh{\left(\frac{\tau hb^2}{2\kT}\right)},
\label{eqvk}
\end{equation}
where we have simplified the general expression (Eq. \ref{eq:kinkWidthEvol})
by considering only migration of $K_+$ kink. 

The dislocation velocity predicted by this analytical approach, 
incorporating the expressions of the nucleation rate $J_{\rm kp}$ (Eq. \ref{eqJTot}) and of the kink-pair extension $L_{\rm kp}$ (Eq. \ref{eq:nucleation_kink_extension}) in Eq. \ref{eqvD}, can be compared to results of kMC simulations. 
As kMC simulations are performed with an idealised energy landscape where the kink-pair formation energy is constant (Fig. \ref{figKPMod}b), the function $f(\tau,T)$ is taken equal to 1 in the expression of the nucleation rate (Eq. \ref{eqJTot}).
Comparison for various temperatures $T$, applied stresses $\tau$, 
and nucleation energy barriers $E^{\rm nuc}$ (Fig. \ref{figvDSimpMod}) shows a very good agreement between both approaches.
In particular, the variation of the dislocation velocity with the dislocation length $L_{\rm d}$ is well reproduced, showing that Eq. \ref{eq:nucleation_kink_extension} manages to describe both the length dependent and the kink collision regimes.
The analytical model also correctly predicts variations of the dislocation velocity with the height of the nucleation barrier $E^{\rm nuc}$: this model is valid whatever the mechanism controlling dislocation mobility, kink-pair nucleation or kink migration.
Going back to the real energy landscape where the formation energy of a kink-pair depends on its width (Fig. \ref{figKPMod}), the same analytical expression of the dislocation velocity can be used but with a function $f(\tau,T)$, which is now different from 1 and needs to be evaluated 
numerically (\cf{} \ref{app:kinkNucleation}).

\subsection{Classical nucleation theory}

Instead of Eq. \ref{eqJTot}, which gives the exact expression of the nucleation rate, one usually relies on the classical nucleation theory (CNT) to predict this nucleation rate.
This approximation replace the discrete summation appearing in the nucleation rate, \ie{} the function $f(\tau,T)$ (see \ref{app:kinkNucleation}), by a continuous integration and evaluate it through an harmonic approximation of the formation enthalpy around the critical size \cite{Clouet2009c}
(see \ref{app:CNT} for the derivation).
The nucleation rate is then given by
\begin{equation}
	J^{\rm CNT}_{\rm kp} = \nu_{\rm D} \, Z(\tau,T) \exp{\left( - \frac{H_{\rm kp}^*(\tau) + E^{\rm mig}_+}{\kT} \right)}
	\label{eqJCNT}
\end{equation}
with $H_{\rm kp}^*(\tau)$ the critical nucleation barrier and $Z(\tau,T)$ the Zeldovitch factor which accounts for fluctuations around the critical size\footnote{In Hirth and Loth model of diffusive glide (chapter 15 in ref. \cite{Hirth1982}), these fluctuations correspond to the factor $b/x'$ in the nucleation rate (Eq. 15-25 in \cite{Hirth1982}).
Using Eq. 15-36 in \cite{Hirth1982} for x', one shows that $b/x' = [(1+\nu)/(1-\nu)]^{1/4}\sqrt{\pi}Z(\tau,T)$, leading to a nucleation rate larger by a factor $\sim 2$ than the one predicted by CNT.}. 
These quantities are defined at the critical size $n^*$, which is the size leading to a maximum of the kink-pair formation enthalpy
\begin{equation*}
	H_{\rm kp}(n,\tau) 
		%= E^{\rm f}_{\rm kp}(nb) - \tau h b^2 n
		= E^{\rm f}_{\rm kp}- \frac{\theta h^2}{n b} - \tau h b^2 n,
\end{equation*}
where we have assumed that the formation energy is given by the elastic interaction model (Eq. \ref{eqFormEnergy}).
This leads to 
\begin{equation*}
	\begin{gathered}
	n^*(\tau) = \sqrt{ \frac{\theta h}{\tau b^3} }, 
	\\
	H_{\rm kp}^*(\tau) = E^{\rm f}_{\rm kp} - 2 \sqrt{\theta h^3 b \tau}, 
	\\
	Z(\tau,T) = \sqrt{ - \frac{1}{2\pi\kT} \left. \frac{\partial^2 H_{\rm kp}}{\partial n^2} \right|_{n=n^*}} 
		= \sqrt{ \frac{1}{\pi\kT} } \ \sqrt[4]{ \frac{ \tau^3 h b^7}{\theta} }.
	\end{gathered}
\end{equation*}

\begin{figure}[bth]
\centering
\includegraphics[width=0.99\linewidth]{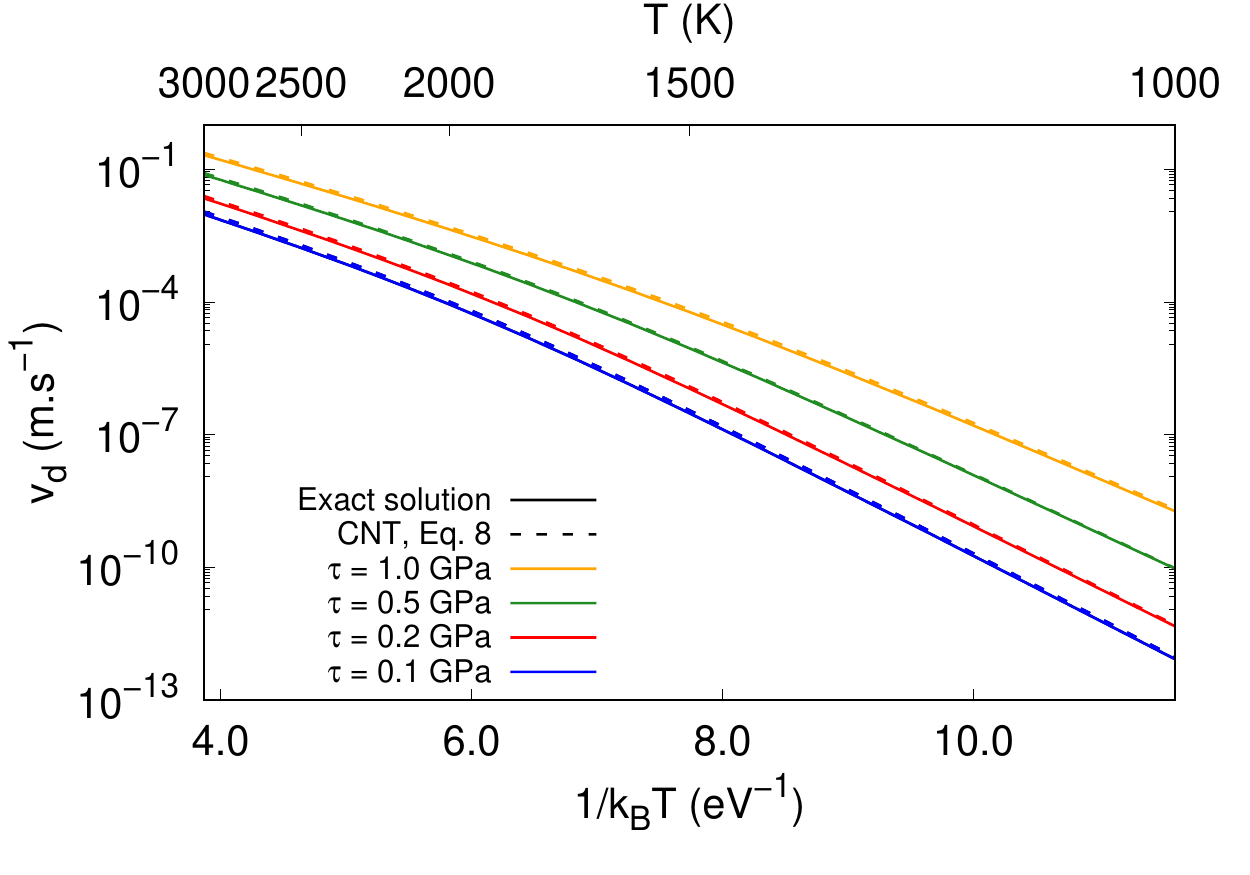}
\caption{Dislocation velocity as a function of the temperature for different external stresses $\tau$ and for a dislocation length $L_{\rm d}=1000\,b\sim275$\,nm with the nucleation rate given by Eq. \ref{eqJTot} (exact solution) and by Eq. \ref{eqJCNT} (CNT).}
\label{figvDlowtau}
\end{figure}

The dislocation velocity obtained using both nucleation rates, Eq. \ref{eqJTot} for the exact solution and Eq. \ref{eqJCNT} for CNT, are presented in Fig. \ref{figvDlowtau} as a function of the temperature for different applied stresses.
In this temperature range where experiments show that screw dislocations experience a lattice friction (section \ref{S22}) and where carbon segregation in the dislocation core is thermodynamically stable \cite{Hachet2020c}, CNT reproduces well the kink-pair nucleation rate, and thus the dislocation velocity. 
The good achievement of CNT confirms that the activation enthalpy controlling dislocation velocity 
is the sum of the kink-pair formation enthalpy and of the kink migration energy,
$E^{\rm f}_{\rm kp} + E^{\rm mig}_{+} - 2 \sqrt{\theta h^3 b \tau}$,
with a slight stress dependence which can be neglected at low stress.
This contrasts with the Peierls regime at low temperature existing in pure BCC metals, 
in particular in tungsten, where kinks can freely glide along the dislocation, 
leading to an activation enthalpy for dislocation velocity which simply equals the kink-pair formation enthalpy
\cite{Brunner2000a,Brunner2010,Clouet2021}.
Because of the covalent bonding between C and W atoms inside the reconstructed core,
similar to the one existing in WC tungsten carbide \cite{Luthi2017}, 
kinks on screw dislocation in presence of carbon are highly localised 
and are difficult to move. As a consequence, the dislocation velocity
in the Peierls regime reappearing at high temperature can be described with the same expression 
as the one which generally applies in covalent systems or semi-conductors \cite{Hirth1982,Caillard2003}.

\subsection{Comparison with experiments}

The dislocation mobility predicted by this kinetic model can be compared with the experimental one.
The experimental velocity is evaluated from the dislocation displacement observed with TEM in Fig. \ref{fig:WC_S1_1150C}. 
We obtain a dislocation velocity of 2\,nm.s$^{-1}$ at 1423\,K.
Using Eqs. (\ref{eqvD}) and (\ref{eqJTot}), the applied stress $\tau$ corresponding to this velocity and temperature should be around 7\,MPa for a dislocation length $L_{\rm d}=275$\,nm.
This result is in the same range as the yield stress measured experimentally in previous work by Taylor \cite{Taylor1965}, using tensile tests on tungsten single crystal with a carbon concentration of 15\,appm: the experimental yield stress varies between 14 and 28\,MPa at 1644\,K.
This demonstrates the ability of our kinetic model parametrised on \abinitio{} calculations to describe dislocation mobility in the temperature range where carbon segregates on screw dislocations.

One can also predict the critical temperature at which 
the yield stress necessary to activate dislocation glide vanishes.
Using Orowan law, the plastic strain rate depends on dislocation velocity through 
$\dot{\varepsilon}=\rho_{\rm d} \, b \, v_{\rm d}$, 
with $\rho_{\rm d}$ the dislocation density.
In the length dependent regime, the average extension of a kink-pair
is equal to the dislocation length and depends on the dislocation density through
$L_{\rm kp}=1/\sqrt{\rho_{\rm d}}$.
Eq. \ref{eqvD} for dislocation velocity thus leads to 
$\dot{\varepsilon} = \sqrt{\rho_{\rm d}} \, h \, J_{\rm kp}$, 
showing that the activation enthalpy is 
$H^{\rm act} = E^{\rm f}_{\rm kp} + E^{\rm mig}_{+} -2\sqrt{\theta h^3 b \tau}$.
The athermal temperature is obtained by considering this activation enthalpy
in the limit of a vanishing stress and by inverting the dependence of the strain rate 
with the temperature, leading to 
\begin{equation}
	T_{\rm a} = \frac{ E^{\rm f}_{\rm kp} + E^{\rm mig}_{+} }
	{k \ln{( \sqrt{\rho_{\rm d}} \nu_{\rm D} h / \dot{\varepsilon} )} },
	\label{eq:Ta}
\end{equation}
where we have neglected the Zeldovitch factor.
Considering a typical strain rate $\dot{\varepsilon}=10^{-3}$\,s$^{-1}$
and a dislocation density $\rho_{\rm d}$ ranging from $10^{8}$ to $10^{12}$\,m$^{-2}$, 
one obtains an athermal temperature between 1500 and 1800\,K. 
When the Zeldovitch factor is considered, one cannot find anymore an analytical expression of the athermal temperature
which needs to be found numerically\footnote{With the Zeldovitch factor, 
the shear stress necessary to activate dislocation glide becomes null only for an infinite temperature. 
The athermal temperature is then defined as the temperature where this stress 
is equal to the backstress given by Taylor relation $\tau = 0.4\,\mu\,b\sqrt{\rho_{\rm d}}$}. 
Full account of fluctuations around the kink-pair critical size through the Zeldovitch factor
leads to higher values, between 1700 and 2700\,K for the same strain rate and dislocation densities.
A lower athermal temperature is obtained with the kink collision regime 
as the activation enthalpy then becomes 
$H^{\rm act} = 1/2\,E^{\rm f}_{\rm kp} + E^{\rm mig}_{+} -2\sqrt{\theta h^3 b \tau}$.
In all cases, except maybe for the very low dislocation densities, this athermal temperature is lower than the temperature at which carbon segregation on screw dislocations drops ($\sim 2500$\,K) \cite{Hachet2020c}, 
thus showing that the stress necessary to make screw dislocation glide with their carbon atoms will vanish before the carbon segregation disappears.
On the other hand, experiments realised by Taylor \cite{Taylor1965} show a higher athermal temperature, around 2700\,K.
Another rate limiting mechanism controlling dislocation mobility at the highest temperatures, 
\ie{} above 1800\,K, appears therefore necessary to explain such a high experimental athermal temperature. 
This mechanism should have an activation energy higher than the sum of the nucleation and migration 
of kink-pairs, $E^{\rm f}_{\rm kp} + E^{\rm mig}_{+}=3.65$\,eV.
Migration of $K_{-}$ kink may have a role to play. 
Although this kink is immobile in the kMC simulations and its migration can be neglected 
in the analytical expression of the dislocation velocity, part of this result is a consequence
of the periodic boundary conditions used to model an infinite screw dislocation 
which can move to the next Peierls valley simply by migration of the $K_{+}$ kink
once a kink-pair has nucleated.  For a screw dislocation of finite length, 
migration of the $K_{-}$ kink is necessary to eliminate the kink-pair 
at one of the two dislocation end-points.  Nevertheless, considering in Eq. \ref{eq:Ta}
migration energy for $K_{-}$ instead of $K_{+}$ leads to athermal temperatures 
between 1900 and 2300\,K, still too low compared to experiments.
A good candidate for a mechanism limiting dislocation velocity at higher temperatures will be the cross-kinks formed when the screw dislocations
start to glide in several planes, leading to the nucleation of kink-pairs 
in different \hkl{110} planes on the same line \cite{Suzuki1979,Argon2007,Rao2017,Zhao2020,Maresca2020,Zhou2021}.

\section{Conclusion}
\label{S7}

\Insitu{} TEM straining experiments performed on high-purity tungsten containing only 1\,appm of carbon
reveal the reappearance of a lattice friction opposing glide of screw dislocations
above 1373\,K, at temperatures high enough to activate carbon diffusion and to enable segregation 
of carbon atoms in the core of screw dislocation, as predicted by \abinitio{} calculations and thermodynamics
\cite{Hachet2020c}.
\Abinitio{} calculations confirm that the Peierls energy barrier opposing glide of a screw dislocation is too high (1.65\,eV/$b$) to allow for athermal glide of screw dislocations when their cores are fully saturated by C atoms.
Screw dislocations have to glide through the formation and propagation of kink-pairs, a Peierls mechanism in agreement with the viscous glide seen experimentally.
Modelling at the atomic scale of kink-pairs show that kinks are very sharp, only $1b$ wide, thus allowing for the determination of their formation, nucleation and migration energies with \abinitio{} calculations. 
Knowing the full energy landscape of a kink-pair on a screw dislocation decorated with carbon atoms, the dislocation velocity is obtained as a function of the temperature and of the applied stress thanks to kinetic Monte Carlo simulations and to an analytical model relying on classical nucleation theory.  
This kinetic model parametrised on \abinitio{} calculations leads to dislocation velocities compatible with TEM observations in the experimental stress range.
The activation energy controlling dislocation velocity is the sum of the kink-pair formation energy and of its migration energy, leading to 3.65\,eV.

\textbf{Acknowledgments} -
The authors thank Dr. B. Lüthi, Prof. D. Rodney, and Dr. F. Willaime for fruitful discussions.
They acknowledge the financial support from the ANR project DeGAS (ANR-16-CE08-0008).
This work was performed using HPC resources from CINECA computer centre within the framework of the EUROfusion Consortium and from GENCI-TGCC and GENCI-IDRIS computer centres under Grant No. A0070906821.
It has been carried out within the
framework of the EUROfusion Consortium and has received funding from the Euratom research and training program 2014-2018
under grant agreement No 633053. The views and opinions expressed herein do not necessarily reflect those of the European
Commission.

\appendix
\section{Kink-pair nucleation rate}
\label{app:kinkNucleation}

To derive the expression of the kink-pairs nucleation rate $J_{\rm kp}$, 
we follow the same approach as in classical nucleation theory \cite{Clouet2009c}
and describe the population of kink-pairs on a dislocation by their size distribution.
We note $C_{n}$ the concentration of a kink-pair with a width $w_{\rm kp}=n\,b$.
The concentration $C_{0}$ is the probability that the dislocation does not contain any kink
and we will make a dilute limit assumption, $C_{0}\sim1$.

We call $\beta_n$ and $\alpha_n$ the growth and decay rates of a kink-pair of size $n$.
For the kinetic model described in section \ref{sec:EKinkedDislo}, 
they are given $\forall n\geq1$ by
\begin{align*}
	\begin{split}
		\beta_n  &= \nu_{\rm D} \left[
		\exp{\left( - \frac{ E^{\rm mig}_{+} }{\kT} \right)} +
		\exp{\left( - \frac{ E^{\rm mig}_{-} }{\kT} \right)}
		\right] \\
	& \qquad \exp{\left( - \frac{E^{\rm f}_{n+1} - E^{\rm f}_n - \tau h b^2}{2\kT} \right)},
	\end{split} 
	\\
	\begin{split}
		\alpha_{n+1}  &= \nu_{\rm D} \left[
		\exp{\left( - \frac{ E^{\rm mig}_{+} }{\kT} \right)} +
		\exp{\left( - \frac{ E^{\rm mig}_{-} }{\kT} \right)}
		\right] \\
	& \qquad \exp{\left( \frac{E^{\rm f}_{n+1} - E^{\rm f}_n - \tau h b^2}{2\kT} \right)},
	\end{split} 
\end{align*}
and for the nucleation and annihilation rates by
\begin{align*}
	\beta_0  &= \nu_{\rm D} \exp{\left( - \frac{ E^{\rm nuc}}{\kT} \right)}
		\exp{\left( \frac{ \tau h b^2}{2\kT} \right)}, \\
		\alpha_1 &= \nu_{\rm D} \exp{\left( - \frac{ E^{\rm nuc} - E^{\rm f}_1 }{\kT} \right)}
		\exp{\left( - \frac{\tau h b^2}{2\kT} \right)}.
\end{align*}

The flux between kink-pairs of size $n$ to $n+1$ is
\begin{equation*}
	J_{n\to n+1} = \beta_n C_n - \alpha_{n+1} C_{n+1} .
\end{equation*}
For $n\geq1$, this flux can be rewritten
\begin{multline*}
	%J_{n\to n+1} = \beta \, \delta 
	%\exp{\left( - \frac{ E^{\rm f}_n + E^{\rm f}_{n+1}  - 2 n \tau h b^2}{2\kT} \right)}
	%\\
	%\left[ C_n  \exp{\left( \frac{ E^{\rm f}_n - n \tau h b^2}{\kT} \right)} 
			%\right. \\  \left.
			%- C_{n+1} \exp{\left(  \frac{ E^{\rm f}_{n+1} 
			%- (n+1) \tau h b^2}{\kT} \right)}
			%\right],
	J_{n\to n+1} = \beta \, \delta^{2n+1} 
	\exp{\left( - \frac{ E^{\rm f}_n + E^{\rm f}_{n+1}}{2\kT} \right)}
	\\
	\left[ \frac{ C_n }{ \delta^{2n} }  \exp{\left( \frac{ E^{\rm f}_n }{\kT} \right)} 
			%\right. \\  \left.
			- \frac{ C_{n+1} }{ \delta^{2(n+1)} } \exp{\left(  \frac{ E^{\rm f}_{n+1} }{\kT} \right)}
			\right],
\end{multline*}
where we have defined the two quantities
\begin{align}
	\beta &= \nu_{\rm D} \left[
		\exp{\left( - \frac{ E^{\rm mig}_{+} }{\kT} \right)} +
		\exp{\left( - \frac{ E^{\rm mig}_{-} }{\kT} \right)}
		\right]
	\label{eq:nucleation_beta}
	\\
	\delta &= \exp{ \left( \frac{\tau h b^2}{2\kT} \right) }.
	\label{eq:nucleation_delta}
\end{align}

At steady state, all fluxes need to be equal: $J_{n\to n+1}=J_{\rm kp}$.
Using the previous expression obtained for the flux $J_{n\to n+1}$,
one can obtain by summation between sizes $n_1$ and $n_2$
\begin{multline}
	\frac{J_{\rm kp}}{\beta} 
	\sum_{n=n_1}^{n_2}{ \frac{1}{\delta^{2n+1}} \exp{\left(  \frac{ E^{\rm f}_n + E^{\rm f}_{n+1} }{2\kT} \right)} }
	\\ 
	= \frac{ C_{n_1} }{ \delta^{2n_1} } \exp{\left( \frac{ E^{\rm f}_{n_1} }{\kT} \right)} 
	- \frac{ C_{n_2+1} }{ \delta^{2(n_2+1)} } \exp{\left( \frac{ E^{\rm f}_{n_2+1} }{\kT} \right)}.
	\label{eq:J_sum}
\end{multline}
We choose $n_1=1$ and $n_2$ large enough so that the last term in the right hand side can be neglected.
This leads to the following expression for the nucleation rate
\begin{equation*}
	J_{\rm kp} \sum_{n=1}^{\infty}{ \frac{1}{\delta^{2n-1}} \exp{\left(  \frac{ E^{\rm f}_n + E^{\rm f}_{n+1}}{2\kT} \right)} }
		= \beta \, C_{1} \exp{\left( \frac{ E^{\rm f}_{1}}{\kT} \right)}.
\end{equation*}
We define the function 
\begin{equation}
	f(\tau,T) =  \sum_{n=1}^{\infty}{ \frac{\delta^2 - 1}{\delta^{2n}} 
	\exp{\left(  \frac{ E^{\rm f}_n + E^{\rm f}_{n+1} - 2E^{\rm f}_{\infty} }{2\kT} \right)} },
	\label{eq:nucleationSizeFunction}
\end{equation}
where $E^{\rm f}_{\infty}$ is the formation energy of two infinitely separated kinks ($n\to\infty$).
This function $f(\tau,T)$ contains all the effect of the variation of the kink-pair formation energy 
with its size.  When the formation energy is constant ($E^{\rm f}_n=E^{\rm f}_{\infty} \forall n$)
as assumed in the kMC simulations of section \ref{sec:KMC}, this function is equal to 1.
With this definition, the steady-state nucleation rate becomes
\begin{equation*}
	J_{\rm kp} = \beta \frac{ \delta^2 - 1 }{  \delta \, f(\tau,T) } C_{1} \exp{\left( \frac{ E^{\rm f}_{1} - E^{\rm f}_{\infty}}{\kT} \right)}.
\end{equation*}
Considering now the nucleation and annihilation of a kink-pair, this nucleation rate can also be written
\begin{equation*}
	J_{\rm kp} = \beta_0 C_0 - \alpha_1 C_1 = \beta_0 - \alpha_1 C_1 .
\end{equation*}
Combining both expressions of the nucleation rate to eliminate $C_1$,
we obtain
\begin{equation*}
	J_{\rm kp} = \frac{ \beta ( \delta - \delta^{-1} ) }
	{ f(\tau, T)\dfrac{\alpha_1}{\beta_0} \exp{\left( - \dfrac{E^{\rm f}_1 - E^{\rm f}_{\infty}}{\kT} \right)} 
	+ \dfrac{\beta}{\beta_0} (\delta - \delta^{-1})},
\end{equation*}
which leads to the final result
\begin{multline}
	J_{\rm kp} = 2 \nu_{\rm D} \left[ \exp{\left( - \frac{ E^{\rm mig}_{+} }{\kT} \right)} + \exp{\left( - \frac{ E^{\rm mig}_{-} }{\kT} \right)} \right]
		\sinh{\left( \frac{\tau h b^2}{2\kT} \right)}
		/ \\
		\left\{ 
		\left[ \exp{\left( \frac{ E^{\rm nuc} - E^{\rm mig}_{+} }{\kT} \right)} + \exp{\left( \frac{ E^{\rm nuc} - E^{\rm mig}_{-} }{\kT} \right)} \right]
		\right.
		\\
		\times \left[ 1 - \exp{\left( - \frac{ \tau h b^2}{\kT} \right)} \right]
		\\
		\left.
		+ f(\tau, T) \exp{\left( \frac{E^{\rm f}_{\infty} - \tau h b^2 }{\kT} \right)}
		\right\}.
		\label{eq:nucleationRate}
\end{multline}

We need now to calculate the function $f(\tau,T)$ defined by Eq. \ref{eq:nucleationSizeFunction}.
For a size $n$ large enough ($n\geq2$ in the present study), the kink-pair formation energy 
can be evaluated with the kink elastic interaction model (Eq. \ref{eqFormEnergy}). 
There is thus a size $n_{\infty}$ large enough for which the argument of the exponential 
appearing in the definition of the function $f(\tau,T)$ is small enough
to make the approximation
\begin{equation*}
	\sum_{n=n_{\infty}}^{\infty}{ \frac{1}{\delta^{2n}} 
	\exp{\left(  \frac{ E^{\rm f}_n + E^{\rm f}_{n+1} - 2E^{\rm f}_{\infty} }{2\kT} \right)} }
	\sim
	\sum_{n=n_{\infty}}^{\infty}{ \frac{1}{\delta^{2n}} \left( 1 - \frac{\theta h^2}{nb\kT} \right)}.
\end{equation*}
We then use the result
\begin{equation*}
	\sum_{n=1}^{\infty}{ \frac{1}{\delta^{2n}} \left( 1 - \frac{\theta h^2}{nb\kT} \right)}
	=
	\frac{1}{\delta^2 -1} + \frac{\theta h^2}{b\kT} \ln{\left( 1-\frac{1}{\delta^2} \right)},
\end{equation*}
to finally obtain
\begin{multline*}
	f(\tau,T) = 1 + \frac{\theta h^2}{b\kT} \left( \delta^2-1 \right) \ln{\left( 1-\frac{1}{\delta^2} \right)}
	\\
	+ \sum_{n=1}^{n_{\infty}}{ \frac{\delta^2 - 1}{\delta^{2n}} 
	\left[
	\exp{\left(  \frac{ E^{\rm f}_n + E^{\rm f}_{n+1} - 2E^{\rm f}_{\infty} }{2\kT} \right)}
	- 1 + \frac{\theta h^2}{nb\kT}
	\right] },
\end{multline*}
with all the stress dependence incorporated in the parameter $\delta$ (Eq. \ref{eq:nucleation_delta}).
This expression can be used to evaluate the function $f(\tau,T)$ by properly choosing
the cutoff $n_{\infty}$ so the term appearing in the sum is small
and the exponential is well evaluated by its series expansion.
This corresponds to the condition $n_{\infty} \gg \theta h^2 / b\kT$, 
thus allowing to restrict the sum to a reasonable number of terms to evaluate it numerically.

\section{Classical nucleation theory}
\label{app:CNT}

The nucleation rate given by classical nucleation theory (CNT) can be recovered starting from Eq. \ref{eq:J_sum}, 
where we now use $n_{1}=0$, thus assuming that the first energy barrier remains smaller than the followings, 
and still $n_2=\infty$.
This equation then becomes
\begin{equation*}
	\frac{J_{\rm kp}}{\beta} 
	\sum_{n=0}^{\infty}{ \frac{1}{\delta^{2n+1}} \exp{\left(  \frac{ E^{\rm f}_n + E^{\rm f}_{n+1} }{2\kT} \right)} }
	= 1,
\end{equation*}
where we have used the property $C_0=1$.
Defining the formation enthalpy $H_{\rm kp}(n,\tau) = E^{\rm f}_n - \tau h b^2 n$, 
one obtains
\begin{equation*}
	\frac{J_{\rm kp}}{\beta} 
	\sum_{n=0}^{\infty}{ \exp{\left(  \frac{ H_{\rm kp}(n,\tau) + H_{\rm kp}(n+1,\tau) }{2\kT} \right)} }
	= 1.
\end{equation*}
The sum appearing in this equation can be evaluated with the usual approach of CNT \cite{Clouet2009c}. 
As only terms around the critical size $n^*$, where the enthalpy is maximum, 
have a non negligible contribution to the sum,
the enthalpy function is replaced by its second order Taylor series around this size 
and the discrete summation is replaced by a continuous integration over all reals
\begin{multline*}
	\sum_{n=0}^{\infty}{ \exp{\left(  \frac{ H_{\rm kp}(n,\tau) + H_{\rm kp}(n+1,\tau) }{2\kT} \right)} }
	\\
	\sim 
	\int_{-\infty}^{\infty}{ \exp{\left(
		\frac{H^*_{\rm kp}}{\kT} 
		+ \frac{H^{''}_{\rm kp}}{4\kT} \left[ (n-n^*)^2 + (n+1-n^*)^2 \right] 
	\right)} \mathrm{d}n },
\end{multline*}
where $H^*_{\rm kp}$ is the value of the enthalpy and $H^{''}_{\rm kp}$ its second derivative 
at the critical size.
This leads to the result
\begin{multline*}
	\sum_{n=0}^{\infty}{ \exp{\left(  \frac{ H_{\rm kp}(n,\tau) + H_{\rm kp}(n+1,\tau) }{2\kT} \right)} }
	\\
	\sim 
	\exp{\left( \frac{H^*_{\rm kp}}{\kT} \right)}
	\exp{\left( \frac{H^{''}_{\rm kp}}{8\kT} \right)} \sqrt{- \frac{2\pi\kT}{H^{''}_{\rm kp}} }.
\end{multline*}
Defining the Zeldovitch factor by
\begin{equation}
	Z(\tau,T) = \sqrt{ - \frac{H^{''}_{\rm kp}}{2\pi\kT} }
	\exp{\left( - \frac{H^{''}_{\rm kp}}{8\kT} \right)},
	\label{eq:Zeldo}
\end{equation}
one obtains the nucleation rate of CNT:
\begin{equation*}
	J_{\rm kp} = \beta \,Z(\tau,T) \exp{\left( -\frac{H^*_{\rm kp}}{\kT} \right)},
\end{equation*}
corresponding to Eq. \ref{eqJCNT} in the main text.
The Zeldovitch factor slightly differs from the usual one because of the exponential appearing in Eq. \ref{eq:Zeldo}.
This is a consequence of our kinetic model where the kink migration energy is obtained from the average 
of the initial and final enthalpy. Nevertheless, one can usually assume that $H^{''} \ll \kT$, 
leading to the known expression
$ Z(\tau,T) = \sqrt{ - H^{''}_{\rm kp} / 2\pi\kT }$.

\section{Time evolution of a kink-pair}
\label{app:kinkVelocity}

To determine how the width of a kink-pair grows with time,
we first compute the kink velocity. 
We discretise a screw dislocation with a $1\,b$ step
and note $p_i$ the probability to find a kink at the position $z=i\,b$.
Considering first a kink, like $K_{-}$ in Fig. \ref{figDisloDK8b}, 
moving in the direction of the applied stress $\tau$ for increasing $z$, 
the flux of kinks between sites $i$ and $i+1$ is
\begin{equation*}
	\phi^{\rm k}_{i \to i+1} = \Gamma^0 \left( p_i \, \delta - p_{i+1} \, \delta^{-1} \right),
\end{equation*}
where we have defined the unbiased jump frequency $\Gamma^0$ 
and the drift contribution $\delta$ by
\begin{equation*}
	\Gamma^0 = \nu_{\rm D} \exp{\left( - \frac{E^{\rm mig}}{\kT} \right)} 
	\textrm{\quad and \quad}
	\delta   = \exp{ \left( \frac{\tau h b^2}{2\kT} \right) }.
\end{equation*}
The time evolution of the kink probability distribution is given by the following master equation
\begin{align*}
	\frac{\partial p_i}{\partial t} &= \phi^{\rm k}_{i-1 \to i} - \phi^{\rm k}_{i \to i+1} \\
	&= \Gamma^0 \left[ p_{i-1} \, \delta - p_i \, (\delta + \delta^{-1}) + p_{i+1} \, \delta^{-1} \right],
\end{align*}
which corresponds to a discretised diffusion equation.

Assuming now that the dislocation contains initially a single kink, 
its average position at a time $t$ is defined by
\begin{equation*}
	\langle z \rangle = b \sum_i {i \, p_i}.
\end{equation*}
The time evolution of the kink average position is thus
\begin{align*}
	\frac{\partial \langle z \rangle}{\partial t} 
	&= b \sum_i{ i \frac{\partial p_i}{\partial t} }, \\
	&= b \, \Gamma^0 \sum_i{ i \left[ p_{i-1} \, \delta - p_i \, (\delta + \delta^{-1}) + p_{i+1} \, \delta^{-1} \right] }, \\
	&= b \, \Gamma^0 \, ( \delta - \delta^{-1}) \sum_i{p_i}.
\end{align*}
As we assume a single kink on the dislocation line and since the master equation describing kink diffusion is conservative,
the sum appearing in the right hand side is equal to 1 at any time $t$.
The average position of the kink is thus
\begin{align*}
	\langle z \rangle &= b \, \Gamma_0 \, ( \delta - \delta^{-1}) \, t, \\
	&= 2 \, b \, \nu_{\rm D}\exp{\left( - \frac{E^{\rm mig}}{\kT} \right)} \sinh{ \left( \frac{\tau h b^2}{2\kT} \right)} \, t.
\end{align*}

Going back now to a kink-pair on a dislocation, the two kinks $K_{-}$ and $K_{+}$
diffuse with different migration energies, $E_{-}^{\rm mig}$ and $E_{+}^{\rm mig}$,
and their coupling with the applied stress is of opposite sign. 
The width of the kink-pair is the distance between these two kinks, 
$w_{\rm kp} = \langle z_{K-} \rangle - \langle z_{K+} \rangle$. 
We define from this distance a kink average velocity $v_{\rm k}=w_{\rm kp}(t)/2t$,
leading to the final result
\begin{multline}
	v_{\rm k} = b \, \nu_{\rm D} 
	\left[ \exp{\left( - \frac{E_{-}^{\rm mig}}{\kT} \right)} + \exp{\left( - \frac{E_{+}^{\rm mig}}{\kT} \right)} \right] \\
		\sinh{ \left( \frac{\tau h b^2}{2\kT} \right)}.
	\label{eq:kinkWidthEvol}
\end{multline}
This expression can be further linearised for not too high stress to obtain a drag coefficient for kink velocity \cite{Po2016}.

%\section*{References}
\bibliographystyle{elsarticle-num-names}
\biboptions{sort&compress}
\bibliography{BibActaMat}

\end{document}